\documentclass[12pt]{article}
\usepackage{graphicx} 
\usepackage[utf8]{inputenc}
\usepackage{mathbbol}
\usepackage{bm}
\usepackage{url}
\usepackage{amsmath}
\usepackage{natbib}
\usepackage[utf8]{inputenc}
\usepackage{xcolor}
\usepackage{graphicx}
\usepackage{enumerate}
\usepackage{amsthm}
\usepackage{amsmath}
\usepackage{amssymb}
\usepackage{marginnote}
\usepackage{mathtools}
\usepackage{float}
\usepackage{framed}
\usepackage{nameref,hyperref}

\usepackage{booktabs} 
\usepackage{changepage}
\usepackage{color}
\usepackage{setspace}
\usepackage{lscape}
\usepackage{algorithm}
\usepackage{algpseudocode}

\usepackage[top=0.7in, bottom=0.7in, left=1in, right=1in]{geometry}

\usepackage{array}
\newcolumntype{+}{!{\vrule width 2pt}}
\newlength\savedwidth

\newcommand\thickhline{\noalign{\global\savedwidth\arrayrulewidth\global\arrayrulewidth 2pt}%
\hline
\noalign{\global\arrayrulewidth\savedwidth}}

\begin{document}
\vspace*{0.2in}

\begin{flushleft}
{\Large
\textbf\newline{Bayesian inference of antibody evolutionary dynamics using multitype branching processes}
}
\newline
\\
Athanasios G. Bakis\textsuperscript{1},
Ashni A. Vora\textsuperscript{2},
Tatsuya Araki\textsuperscript{2},
Tongqiu Jia\textsuperscript{3},
Jared G. Galloway\textsuperscript{4},
Chris Jennings-Shaffer\textsuperscript{4},
Gabriel D. Victora\textsuperscript{2,5},
Yun S. Song\textsuperscript{6},
William S. DeWitt\textsuperscript{3,4*},
Frederick A. Matsen IV\textsuperscript{4,3,7,5*},
Volodymyr M. Minin\textsuperscript{1*}
\\
\bigskip
\textbf{1} Department of Statistics, University of California, Irvine, CA, USA
\\
\textbf{2} Laboratory of Lymphocyte Dynamics, The Rockefeller University, New York, NY, USA
\\
\textbf{3} Department of Genome Sciences, University of Washington, Seattle, WA, USA
\\
\textbf{4} Computational Biology Program, Fred Hutchinson Cancer Research Center, Seattle, WA, USA
\\
\textbf{5} Howard Hughes Medical Institute, Seattle, WA, USA
\\
\textbf{6} Department of Electrical Engineering \& Computer Sciences, University of California, Berkeley, CA, USA
\\
\textbf{7} Department of Statistics, University of Washington, Seattle, WA, USA
\bigskip

* co-corresponding authors (wsdewitt@uw.edu, matsen@fredhutch.org, vminin@uci.edu)

\end{flushleft}

\section*{Abstract}
When our immune system encounters foreign antigens (i.e., from pathogens), the B cells that produce our antibodies undergo a cyclic process of proliferation, mutation, and selection, improving their ability to bind to the specific antigen.
Immunologists have recently developed powerful experimental techniques to investigate this process in mouse models.
In one such experiment, mice are engineered with a monoclonal B-cell precursor and immunized with a model antigen.
B cells are sampled from sacrificed mice after the immune response has progressed, and the mutated genetic loci encoding antibodies are sequenced.
This experiment allows parallel replay of antibody evolution, but produces data at only one time point; we are unable to observe the evolutionary trajectories that lead to optimized antibody affinity in each mouse.
To address this, we model antibody evolution as a multitype branching process and integrate over unobserved histories conditioned on phylogenetic signal in sequence data, leveraging parallel experimental replays for parameter inference.
We infer the functional relationship between B-cell fitness and antigen binding affinity in a Bayesian framework, equipped with an efficient likelihood calculation algorithm and Markov chain Monte Carlo posterior approximation.
In a simulation study, we demonstrate that a sigmoidal relationship between fitness and binding affinity can be recovered from realizations of the branching process.
We then perform inference for experimental data from 52 replayed B-cell lineages sampled 15 days after immunization, yielding a total of 3,758 sampled B cells.
The recovered sigmoidal curve indicates that the fitness of high-affinity B cells is over six times larger than that of low-affinity B cells, with a sharp transition from low to high fitness values as affinity increases.

\section*{Introduction}

\subsection*{Scientific background and problem formulation}

Understanding of the adaptive immune system is critical for developing new strategies to fight disease (for example, in immunotherapy and vaccine development) \cite{vollmers2013genetic}.
The antibody-mediated component of the adaptive immune system is carried out by B cells, which produce antibodies.
When antibodies are bound to the membrane of a B cell, they are called B cell receptors (BCRs).
The BCR is determined by its genetic sequence encoded in the immunoglobulin heavy and light chain loci.
An \emph{antigen} is a foreign molecule that is the target of the adaptive immune system; it could be a part of a pathogen or a vaccine.
BCRs improve their binding ability to a given antigen during \textit{affinity maturation}, a process that happens in lymph node structures called \emph{germinal centers}.

What mechanistic details do we currently know about affinity maturation?
During affinity maturation, B cells compete to optimize the antibody supply for a particular target in a cyclic two-step process, alternating between the so-called \emph{light zone} and \emph{dark zone} of a germinal center \cite{victora2022germinal}.
\begin{enumerate}
	\item \textbf{Selection}: B cells compete via their BCRs in the light zone to bind to the antigen; the binding \textit{affinity} is a quantitative measure of binding ability defined in terms of a chemical dissociation constant $K_D$ (having units of molar concentration, with lower values corresponding to higher affinity).
	B cells that succeed in binding antigen present antigen-derived peptides to and receive signals from T cells that trigger them to migrate to the dark zone and initiate proliferation.
	\item \textbf{Mutation}: B cells proliferate in the dark zone, mutating the immunoglobin loci along the way to create new B cells with new, potentially improved BCRs. These cells then migrate back to the light zone.
\end{enumerate}
By some estimates, the rate of mutation in the immunoglobin loci is six orders of magnitude larger than the genome-wide rate \cite{kleinstein2003estimating, scally2012revising}.
Consequently, the immune response sees substantial B cell evolution in only a few weeks.

There is a distinction, however, between the affinity of a B cell's BCRs, and its fitness (cellular proliferation rate) in the affinity maturation process.
BCR affinity is a molecular phenotype that can be measured experimentally outside the context of the germinal center.
However, fitness---thought to be a function of binding affinity---cannot easily be measured directly in the germinal center.
The functional relationship between binding affinity and fitness is therefore not well-characterized quantitatively.
For example, there may exist a threshold of binding affinity that is sufficient for proliferative signals from T cells, and any B cells whose BCRs exceed this threshold have the same fitness \cite{batista1998affinity}.
Thus, in this paper, we pose the question: what is the functional relationship between the binding affinity of an antibody and the fitness of the B cell that produces it?

Current experiments cannot directly interrogate the extent to which binding affinity determines B cell fitness, as this would require longitudinal sampling of germinal center B cells; current technology can only sample cells after sacrificing the model organism.
We opt to develop a likelihood-based statistical method equipped with the ability to handle the unobservable quantities governing affinity maturation, allowing us to learn about this fundamental affinity-fitness relationship without needing to measure it directly.
We apply our methodology to a novel experimental dataset of cross-sectional B cell samples from individual germinal centers.

The recent work of DeWitt, Vora, Taraki et al. (2025) provides us with this experimental data \cite{dewitt2025replaying}.
In a highly controlled system, engineered monoclonal B cells specific for the model antigen chicken gammaglobulin (chIgY) were allowed to affinity mature in response to immunization with their cognate antigen.
Individual germinal centers from several mice were micro-dissected, and immunoglobin loci were sequenced.

It is excitingly rare to observe numerous parallel replications of an evolutionary process, as macroevolutionary studies can only observe one replication of most evolutionary events of interest.
As stated earlier, longitudinal observation of a germinal center is currently infeasible, as sampling cells is destructive to the germinal center (and the mouse) and terminates affinity maturation; thus, this experiment is cross-sectional in nature.
However, we show that, under certain assumptions, control of the germinal center's initial state is sufficient to infer the affinity-fitness relationship using only a single time point from each replicate.

\subsection*{Related work} \label{sec:related-work}

From a mathematical modeling point of view, continuous-time branching processes are a classic model for cell division and form a natural basis for our work.
These models are parameterized in terms of the rates at which lineages are born or die, and allow us to model the unobserved intermediate evolutionary history of the B cells.
We can interpret the rate of birth of a lineage as the fitness of the cell it represents.

Likelihood-based inference for partially-observed branching processes has a long history.
Nee et al. (1994) describe how to use a branching process to infer evolutionary dynamics when lineages can go extinct and become unobservable, a key component of our scientific problem \cite{nee1994reconstructed}.
More recent literature builds upon this work, introducing branching processes that allow for lineages of different discrete character states---observed or marginalized over---to have different evolutionary dynamics \cite{maddison2007estimating, fitzjohn2012diversitree, kuhnert2016phylodynamics, beaulieu2016detecting}.
These models form the foundation of our approach to inferring how properties of B cells (such as BCR binding affinity) map to their birth rate.
FitzJohn (2012) furthers this work to allow for a continuous-valued state, evolving according to a diffusion process \cite{fitzjohn2010quantitative}.
Stadler (2013) and MacPherson et al. (2022) provide further reviews of the developments in this literature \cite{stadler2013recovering, macpherson2022unifying}.
Notably, the permission of transitions between states in these models creates analytical intractabilities when deriving the likelihood; numerical methods are consequently used.
However, Barido-Sottani et al. (2020) derive an analytical approximation to the multitype branching process likelihood that assumes character states don't change in unsampled lineages \cite{barido2020multitype}.

These methods, as they stand, are not equipped to handle the details of our scientific question.
First, we are interested in inferring the parameters of a function relating binding affinity to birth rate.
Only FitzJohn (2012) of the aforementioned literature discusses the specification of flexible functional relationships between rates and states, but its use of diffusion processes to model state changes is difficult to integrate with our prior knowledge of B cell mutation rates.
Second, our experimental setup provides us with a sample of independent replicates of evolution (across different germinal centers), which requires a careful investigation of how to create a conditional branching process likelihood suitable for our sampling regime.
Finally, we will show the analytical approximation of Barido-Sottani et al. (2020) does not hold in our context and introduces biases in inference.

\subsection*{Our contribution}

In our work, we extend the current state-of-the art multitype branching processes by:
(1) inferring parameters of a functional relationship between birth rates and lineage attributes (a trait or phenotype assigned to each type);
(2) accommodating multiple independent, partially observed realizations of the branching process;
(3) demonstrating that conditioning on the observation of at least one lineage is critical for producing valid statistical inference when analyzing multiple realizations of the process; and
(4) allowing for a general parameterization of the rates of type change.
We provide an efficient implementation of our inferential pipeline, without approximations of the branching process likelihood.
We show that the relationship between birth rates and lineage attributes can be recovered in simulation from independent and identically distributed realizations of the branching process, and that some model misspecification still yields partially robust inference.
We demonstrate our method's application to B cell evolution from a real experiment.

\section*{Methods}

\subsection*{Data and experimental setup} \label{sec:setup}

\begin{figure}
	\centering
	\begin{framed}
	\includegraphics[width=1\linewidth]{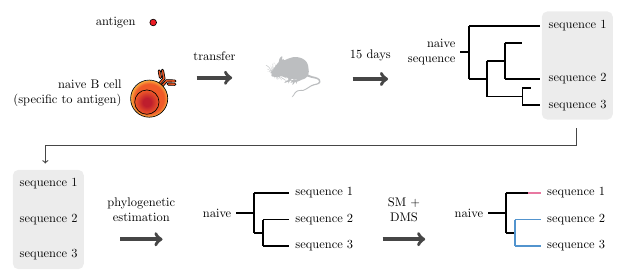}
	\end{framed}
	\caption{
		The data-generating experiment.
		\textbf{Top row:}
		An antigen and B cells are transferred to host mice, initializing affinity maturation.
		While each germinal center in each mouse experiences a full evolutionary process (represented by the tree), we are unable to observe anything before the present (sampling) time.
		The sequences at the tips of the tree are all that is collected as data (shaded box).
		\textbf{Bottom row:}
		Given observed B cell sequences, estimate a phylogenetic tree annotated with mutation times.
		We then are in the unique position to predict the binding affinity of the antibodies produced by any B cell at any point along the tree, using stochastic mapping (SM) and deep mutational scanning (DMS) data.
		These inferred and affinity-annotated trees form the input to our model.
	}
	\label{fig:experimental-setup}
\end{figure}

Consider the following experiment.
Mice receive an adoptive transfer of genetically identical monoclonal B cells and are then immunized with the antigen these cells target.
The genetic sequence common to these B cells is termed the \textit{naive sequence}.
Host mice are engineered to be unable to produce germinal centers with their own B cells; thus, the adoptively transferred B cells are the only ones involved in affinity maturation.
After a fixed duration of time (on the order of days), B cells are sampled from germinal centers in the mice, and their immunoglobulin genes are sequenced; sequences are then aligned.
This experiment is illustrated in Fig \ref{fig:experimental-setup}.

With these sequences, we have observed the results of $I$ parallel replications of affinity maturation, one for each germinal center sampled across the mice.
Let $G_{ij}$ be the genetic sequence of the $j^{th}$ B cell sampled from the $i^{\text{th}}$ germinal center ($i = 1, \dots, I$), and $\mathbf{G}_i$ be collection of all these sequences for germinal center $i$.

We are then provided with a predictive model mapping these B cells' genetic sequences to the binding affinities of their antibodies to the antigen, using a technology called Tite-Seq \cite{adams2016measuring}.
Specifically, a library of all single-amino-acid mutations to the naive sequence was constructed, and the change in the log equilibrium dissociation constant ($-\Delta \log_{10} K_D$) of the resulting antibodies were measured.
These are then extended to multi-amino-acid mutations using an additive model \cite{dewitt2025replaying}.
Let $A(G_{ij})$ be the $-\Delta \log_{10} K_D$ prediction for the antibodies produced by a B cell with sequence $G_{ij}$, representing a log fold change between binding affinity of $G_{ij}$ relative to the naive sequence.
Abusing notation slightly, we will call this measure of binding affinity by the term \textit{binding affinity} itself, with higher values indicating stronger binding to the antigen.

\subsection*{Problem formulation} \label{sec:problem}
Our scientific goal is to infer a mapping from the affinity of a B cell's antibodies (collected in a set $\mathcal{X}$) to its fitness in the germinal center.
Mathematically, we infer a mapping $\lambda_{\phi} : \mathcal{X} \to \mathbb{R}^+$ (parameterized by $\boldsymbol{\phi}$) from a B cell's antibody binding affinity to a $\lambda_\phi(A(G_{ij}))$ which describes fitness (this will be made more precise in the next section).
Let us emphasize that there is a fair amount of unobserved data in our experiment, which makes this task challenging.
Only a subset of cells alive at the cell collection time are sampled and sequenced; we lack knowledge of any cells that died before collection time or were not sampled.
Additionally, the evolutionary history of these cells (sampled or not) is unknown to us.

\begin{figure}
	\small \centering
	\includegraphics[width=1\textwidth]{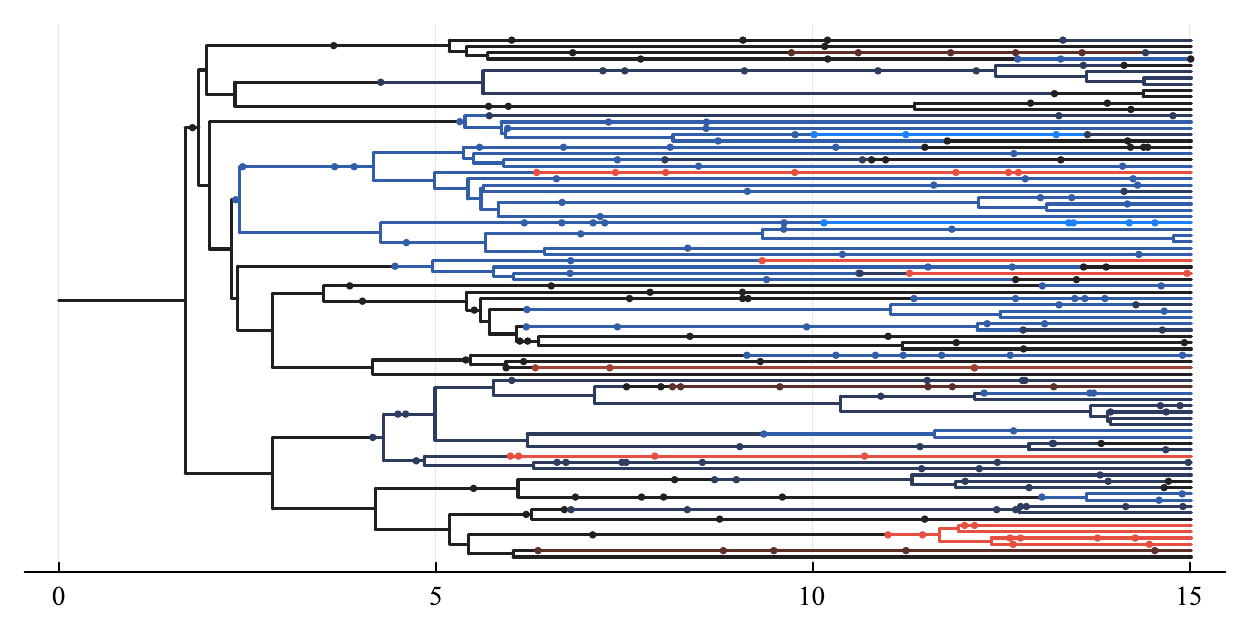}
	\includegraphics[width=1\textwidth]{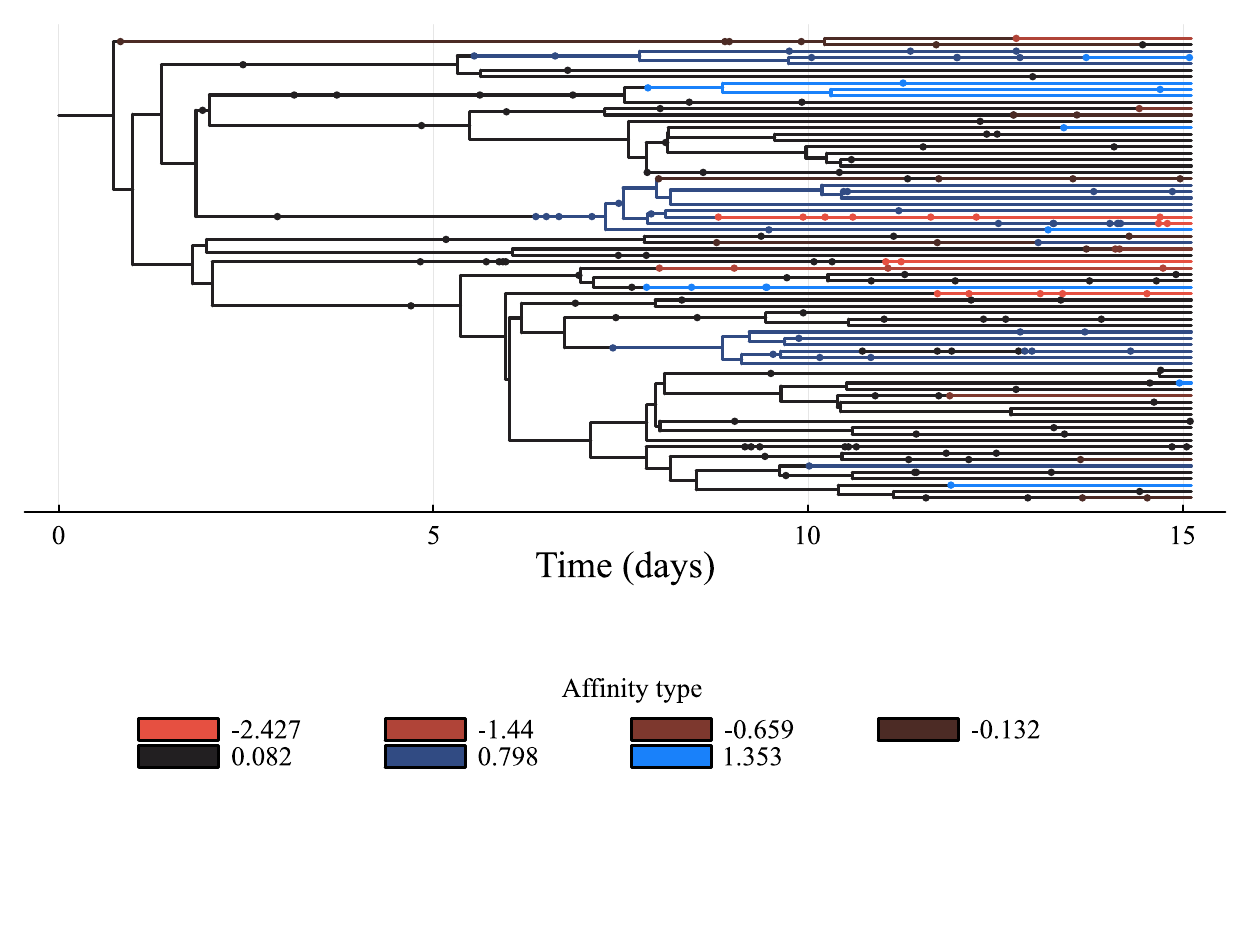}

	\caption{
		One sampled tree from each of two different germinal centers, as output by BEAST.
		Branches are colored by discretized affinity value, and dots represent nucleotide-level B cell mutations (including those that do not change the affinity of the antibody).
	}
	\label{fig:example-beast-trees}
\end{figure}

We will assume, for now, that for each germinal center $i$, we can use $\mathbf{G}_i$ to estimate $T_i$, a phylogenetic tree annotated with $A(G)$ for every sequence $G$ at any point on the tree.
Example trees observed in this manner are shown in Fig \ref{fig:example-beast-trees}.
The branch lengths of the tree are measured in units of clock time.
The relationship between the number of genetic mutations along a branch and the time elapsed is estimable because of our knowledge of the state of the tree at two distinct time points---the root and tips of the tree \cite{drummond2003measurably, duchene2020bayesian}.

\subsection*{Branching process} \label{sec:tree-model}

\begin{figure}
	\centering
	\begin{framed}
	\includegraphics[width=1\linewidth]{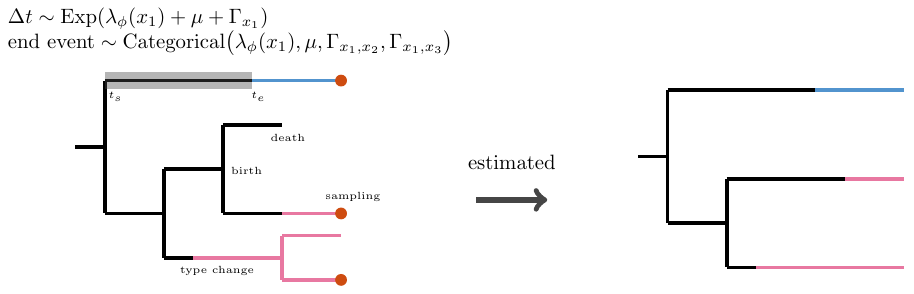}
	\end{framed}
	\caption{
		The data-generating process for the branching process.
		The parameters $\lambda(\cdot), \mu, \Gamma_{\cdot}$ correspond to birth, death, and type change rates, respectively.
		\textbf{Left:}
		Waiting times and events are sampled to create a tree.
		Categorical distribution probabilities are obtained by normalizing the stated weights, which are written as such in the diagram for brevity.
		For example, the probability that a segment in type $x_1$ ends in a birth event is $\lambda_\phi(x_1) / \left[ \lambda_\phi(x_1) + \mu + \Gamma_{x_1, x_2} + \Gamma_{x_1, x_3} \right]$.
		Colors represent types; here, black is type $x_1$, red is type $x_2$, and blue is type $x_3$.
		Lineages which are sampled at collection time are marked with dots on the tips of the tree.
		The highlighted branch segment is explained by the mathematical captions adjacent to them.
		\textbf{Right:}
		Information about lineages that died or were not sampled is lost, yielding the tree on observed lineages.
	}
	\label{fig:data-generating-process}
\end{figure}

We now specify the probability model that relates the phylogenetic tree $T_i$ to our notion of fitness $\lambda_{\phi}$.
Let us first describe the generating process for the tree.
We employ a branching process in which individuals of different ``types'' have different evolutionary dynamics, adopting notation from Barido-Sottani et al. (2020).
Assume there is a finite set $\mathcal{X}$ of possible discrete character states (further referred to as \textit{types}), one of which describes an individual at any given point in time.
The process begins with a single individual of type $x \in \mathcal{X}$ at time $t = 0$.
A continuous-time Markov process is then run, with parameters describing the rates at which the individual either: (1) gives birth to a new individual of the same type, (2) dies, or (3) changes to a different type.
Finally, leaves on the tree are marked as sampled according to specified probability parameters (one for survivors, and one for dead lineages), and subtrees without any sampled lineages are pruned away and considered unobserved.
Fig \ref{fig:data-generating-process} illustrates this process.
Note that type changes do not occur during birth events, and that there is a separate constant birth rate parameter associated with each type.

This model with type-specific birth rates appears useful to us at first glance, as we can use the birth rates as a proxy for the fitness function $\lambda_\phi$ we aim to infer.
In this setup, the possible types $x \in \mathcal{X}$ are the binding affinities of the antibodies produced by the B cells (and thus, $\mathcal{X}$ is an ordered set).
However, the discrete birth rate setup can become problematic as the size of the type space increases, and in our context, there are hundreds of binding affinities to which fitness values must be assigned.
Some existing literature has instead opted to create a continuous type space for the multitype branching process, but models the type change process in a way that is difficult to harmonize with our prior model of B cell mutations \cite{fitzjohn2010quantitative}.
We would like to instead constrain our parameterization in the discrete setup and facilitate information sharing, as well as enable interpolation of the birth rate at types which may not explicitly be in the type space.

Therefore, we build on this notion of type-specific birth rates by replacing the discrete set of birth rates with our choice of parametric curve $\lambda_{\phi} : \mathcal{X} \to \mathbb{R}^+$, e.g., a sigmoidal $\lambda_\phi(x) = \phi_1 / \left[ 1 + \exp(-\phi_2 (x - \phi_3)) \right] + \phi_4$.
Now, our branching process still requires that the type space be finite, so we must discretize the range of possible binding affinities to create this type space.
We thus still only infer $|\mathcal{X}|$ (the cardinality of the set) birth rates in the branching process, but we constrain them to obey the functional form of $\lambda_\phi$.
Our discretization strategy is explained in Section \textit{\nameref{sec:discretization}}.

Notice how the tree realized from this continuous-time Markov process contains more than just the timings of birth events typically seen in a phylogenetic tree.
First, there are death events, marking the end of a lineage prior to collection time.
In our experimental setup, we are unable to sample a dead cell from the germinal center, and thus are always working with trees whose leaves are exclusively survived cells.
Our branching process will accommodate this by setting the sampling probability of dead lineages to zero.
The tree now also contains type change events, where a cell of one type changes to another.
Our types are discretized binding affinities, which we assume we can measure.

Let us now formally derive the density calculation of our branching process, in the spirit of Barido-Sottani et al. (2020).
Recall that every point on the tree is annotated with a binding affinity value $A(\cdot)$.
Let $\mathcal{X}$, the type space, be a finite set of numbers corresponding to discretized binding affinity values, for some discretization function.
Define the following parameters, which will collectively be referred to as $\boldsymbol{\theta}$:
\begin{itemize}
	\item $\boldsymbol{\phi}$, the parameters which define the birth rate function $\lambda_{\phi} : \mathcal{X} \to \mathbb{R}^+$;
	\item $\mu > 0$, the death rate parameter, chosen to be constant over $\mathcal{X}$ \cite{mayer2017microanatomic};
	\item $\boldsymbol{\Gamma}$, the type change rate matrix of dimension $|\mathcal{X}| \times |\mathcal{X}|$;
		\subitem Let $\Gamma_{x, x'}$ denote the rate at which a cell of type $x$ changes to type $x'$ for $x \neq x'$.
		\subitem Let $-\Gamma_{x, x} \overset{\text{def}}{=} \Gamma_x \overset{\text{def}}{=} \sum_{x' \neq x}\Gamma_{x, x'}$ denote the rate at which a cell of type $x$ changes to any other type.
	\item $\rho$, the probability that a cell alive at the collection time is sampled, and thus observed in the tree. (Recall not every cell present in the germinal center will be sampled.)
\end{itemize}
Note that Barido-Sottani et al. (2020) include a parameter $\sigma$ for the probability that a dead lineage is sampled for inclusion in the tree.
As discussed earlier in this section, we set $\sigma = 0$ to align the branching process with our experimental setup of sampling only living cells.

Our goal is to compute the density $p(T_i | \boldsymbol{\theta})$ for the tree $T_i$.
Recall $T_i$ contains information of both the births and type change events of sampled B cell lineages in germinal center $i$; our strategy for annotating type changes is discussed in Section \textit{\nameref{sec:data-description}}.
Note that we measure the times $t > 0$ of these events as the duration of time remaining until collection time $0$.

Let us first define an auxiliary probability that is useful in computing this density.
Let $p_x(t)$ be the probability that a cell with type $x$ at time $t > 0$ is not observed at collection time $0$ in the tree, nor are any of its descendants.
Following Barido-Sottani et al. (2020), $p_x$ is defined through the following system of ordinary differential equations (over $x \in \mathcal{X}$):
\begin{equation} \label{eq:px}
\begin{aligned}
p_x(0) &= 1-\rho, \\
\frac{dp_x}{dt}(t) &= -(\lambda_\phi(x) + \mu + \Gamma_x) p_x(t) + \mu + \lambda_\phi(x) p_x(t)^2 + \sum_{x' \neq x} \Gamma_{x, x'} p_{x'}(t).
\end{aligned}
\end{equation}

To assign probability density to the entire tree $T_i$, we consider how we can aggregate densities of every \textit{segment} $N$ of $T_i$.
A segment is defined as the stretch of tree branch between any two adjacent events: birth, death, or type change.
Let $t_{s, N}$ and $t_{e, N} < t_{s, N}$ be the start and end times of branch segment $N$, respectively.
We then let $q_N(t), \, t \in [t_{e, N}, t_{s, N}]$ be the probability density of the subtree of $T$ beginning with (and including) the portion of tree segment $N$ occurring after time $t$.
As in Barido-Sottani et al. (2020), this quantity is also defined through an ordinary differential equation,
\begin{equation}
\begin{aligned}
\frac{dq_N}{dt}(t) &= -\left(\lambda_\phi(x_N) + \mu + \Gamma_x \right) q_N(t) + 2 \lambda_\phi(x_N) q_N(t) p_{x_N}(t), \quad \text{with initial condition} \\
q_N(t_{e, N}) &=
	\begin{cases}
	\rho & \parbox{0.5\linewidth}{if $N$ ends in a sampling event at collection time $t_{e, N}=0$,} \\ \\
	\lambda_\phi(x_N) q_{N'}(t_{s, N'})q_{N''}(t_{s, N''}) & \parbox{0.5\linewidth}{if $N$ ends in a birth event at time $t_{e, N} > 0$ yielding child segments $N'$ and $N''$,} \\ \\
	\Gamma_{x_N, x_{N'}} q_{N'}(t_{s, N'}) & \parbox{0.5\linewidth}{if $N$ ends in a type change event at time $t_{e, N} > 0$ yielding segment $N'$,}
	\end{cases}
\end{aligned}
\end{equation}
where $x_N$ is the type of the cell represented by segment $N$.
Notice how the initial conditions are explicitly defined for the segments that end at leaves in the tree (i.e. sampling events), whereas the initial conditions for the other segments require the values of $q$ of their child segments.
By performing a postorder traversal through the tree and computing $q$ during this traversal, we visit every segment in such an order that child segments always have $q$ computed before visiting a parent segment.
This traversal ultimately ends at the root segment $N_\text{root}$, which begins at time $t_{s, N_\text{root}}$.
By construction, the value of $q$ for this root segment is the probability density of the entire tree $T_i$ on observed lineages, i.e.,
\begin{equation} \label{eq:tree-model-unconditioned}
	p(T_i | \boldsymbol{\theta}) = q_{N_\text{root}}(t_{s, N_\text{root}}) .
\end{equation}
Note that exploiting the linearity of the ODE may result in a more efficient implementation; we will consider this in future work.

We have not finished, though, due to the importance of considering a conditional distribution over trees.
With the multitype branching process-induced density in Eq (\ref{eq:tree-model-unconditioned}), any tree $T_i$ is a valid realization of the process---including those where all cells go extinct.
However, in our scientific problem, germinal centers are only sampled if they exist at collection time; our experimental setup has no way of inferring a tree without lineages that survived to the sampling time.
If we do not condition on this information, our death rate parameter will be severely underestimated, as it will appear that cells are rarely dying.
Stadler (2013) explores conditioned branching process likelihoods further (in the context of the constant rate parameters) and indeed makes the recommendation to condition the model on the survival of at least one sampled lineage \cite{stadler2013can}.
However, this conditioning is not common practice when working with multitype branching processes, possibly because the difference between conditioning and not conditioning is not very pronounced when analyzing only one realization of the process \cite{barido2020multitype}.
We will show in the Results that conditioning is important when analyzing multiple trees jointly.

Thus, we condition Eq (\ref{eq:tree-model-unconditioned}) on the event that the tree $T_i$ produced at least one observed lineage.
Fortunately, the probability of this event is easily accessible as a solution of the ODE defined earlier (Eq (\ref{eq:px})); it is $p_{x_\text{root}}(t_{s, N_\text{root}})$, where $x_\text{root}$ is the type of the cell at the root of the tree.
We ultimately work with the conditional tree density, denoted $p_c$:
\begin{equation} \label{eq:tree-model}
p_c(T_i | \boldsymbol{\theta}) = \frac{q_{N_\text{root}}(t_{s, N_\text{root}})}{1-p_{x_\text{root}}(t_{s, N_\text{root}})} .
\end{equation}

\subsection*{Putting it all together} \label{sec:alltogether}

With $\boldsymbol{\theta}$ defined as in Section \textit{\nameref{sec:tree-model}}, our target of inference is:
\begin{equation} \label{eq:target}
p(\boldsymbol{\theta} | \mathbf{T}) \propto p(\boldsymbol{\theta}) \prod_i  p_c(T_i | \boldsymbol{\theta}) .
\end{equation}
Hamiltonian Monte Carlo or any other MCMC, in principle, can be used for this posterior.
There are several considerations to make when adapting this methodology to an experiment, though:
\begin{enumerate}
	\item The functional form of the birth rate $\lambda_{\phi} : \mathcal{X} \to \mathbb{R}^+$ needs to be chosen.
	\item As described in Section \textit{\nameref{sec:tree-model}}, the interval of possible binding affinities needs to be discretized in some manner to create a type space for the branching process.
	\item As discussed by Stadler and Bonhoeffer (2013), there are identifiability concerns around the parameters $\lambda_\phi(\cdot), \mu, \rho$ \cite{stadler2013uncovering}. One solution is to fix one of the parameters (often $\rho$) using prior information. Alternatively, we hypothesize that setting a highly informative prior distribution on one of the parameters is sufficient.
	\item The matrix $\boldsymbol{\Gamma}$ can be inferred in principle, but given prior information about B cell mutation dynamics, it may be preferable to fix this parameter and greatly reduce the overall number of inferred quantities. One approach is to replace $\boldsymbol{\Gamma}$ in the branching process with the term $\delta \boldsymbol{\Gamma}^*$, where $\boldsymbol{\Gamma}^*$ is the known prior-informed rate matrix, and $\delta$ is an unknown scalar parameter to be inferred jointly with other model parameters.
\end{enumerate}
We discuss our own decisions for these points (in relation to our experimental data) in Section \textit{\nameref{sec:adapting}}.

\subsection*{Our implementation}

Bayesian inference was conducted via Hamiltonian Monte Carlo with No-U-Turn Sampling in the Julia programming language using the Turing probabilistic programming library \cite{bezanson2017julia, hoffman2014no, turing}.
Our package implementing a tree simulator and density computation for branching process can be found at \url{https://github.com/thanasibakis/gcdyn.jl}.
The numerical integration for density computation was conducted using the DifferentialEquations.jl package \cite{rackauckas2017differentialequations}.
Our package implementing additional tree simulators and B cell mutation models (for examining robustness to model misspecifications) can be found at \url{https://github.com/matsengrp/gcdyn}.
Our code for the following data analyses and simulation studies can be found at \url{https://github.com/thanasibakis/gcdyn-analyses}.

\section*{Results}

\subsection*{Adapting our methodology to the experimental setup of DeWitt, Vora, Taraki et al. (2025)} \label{sec:adapting}

We now address our modeling decisions for the considerations discussed in Section \textit{\nameref{sec:alltogether}}.

\begin{enumerate}
	\item The birth rate function is specified to have a sigmoidal form,
	\[ \lambda_\phi(x) = \phi_1 / \left \{ 1 + \exp \left [ -\phi_2(x - \phi_3) \right ] \right \} + \phi_4 . \]
	\item We examine genetic sequences from a separate experiment to form our type space; see Section \textit{\nameref{sec:discretization}} below.
	\item We use prior knowledge to fix $\rho$. Specifically, in simulation studies, we assume $\rho$ is known (unless otherwise stated), and in data analysis, we fix individual germinal center-specific parameters estimated as $\rho_i = \frac{|\mathbf{G}_i|}{1000}$, i.e., the observed number of cells in the germinal center divided by the prior-informed total number of cells at time of sampling.
	\item We use prior knowledge to fix $\boldsymbol{\Gamma}$ and introduce a scalar $\delta$ alongside it; see Section \textit{\nameref{sec:tc-matrix}} below.
\end{enumerate}
Thus, the parameters we infer in Eq \ref{eq:target} are $\boldsymbol{\theta} = (\boldsymbol{\phi}, \mu, \delta)$.
The details of some of these decisions here are given below.

\subsubsection*{Discretization of affinity values for type space} \label{sec:discretization}

We now describe how we discretize the continuous range of affinity values we observe in our tree samples.
Naively, we may propose to collect all observed affinity values in both the leaf and ancestral tree nodes across the 52 germinal centers, and create $n$ bins from evenly-spaced cutoff values.
We can then use the median affinity within the bin to be the representative type value for that bin.
However, there may be a bias with this technique; we may observe fewer instances of small binding affinities, since the B cells producing those antibodies would likely not be selected for in the affinity maturation process and thus would not survive.

\begin{figure}
	\includegraphics[width=1\textwidth]{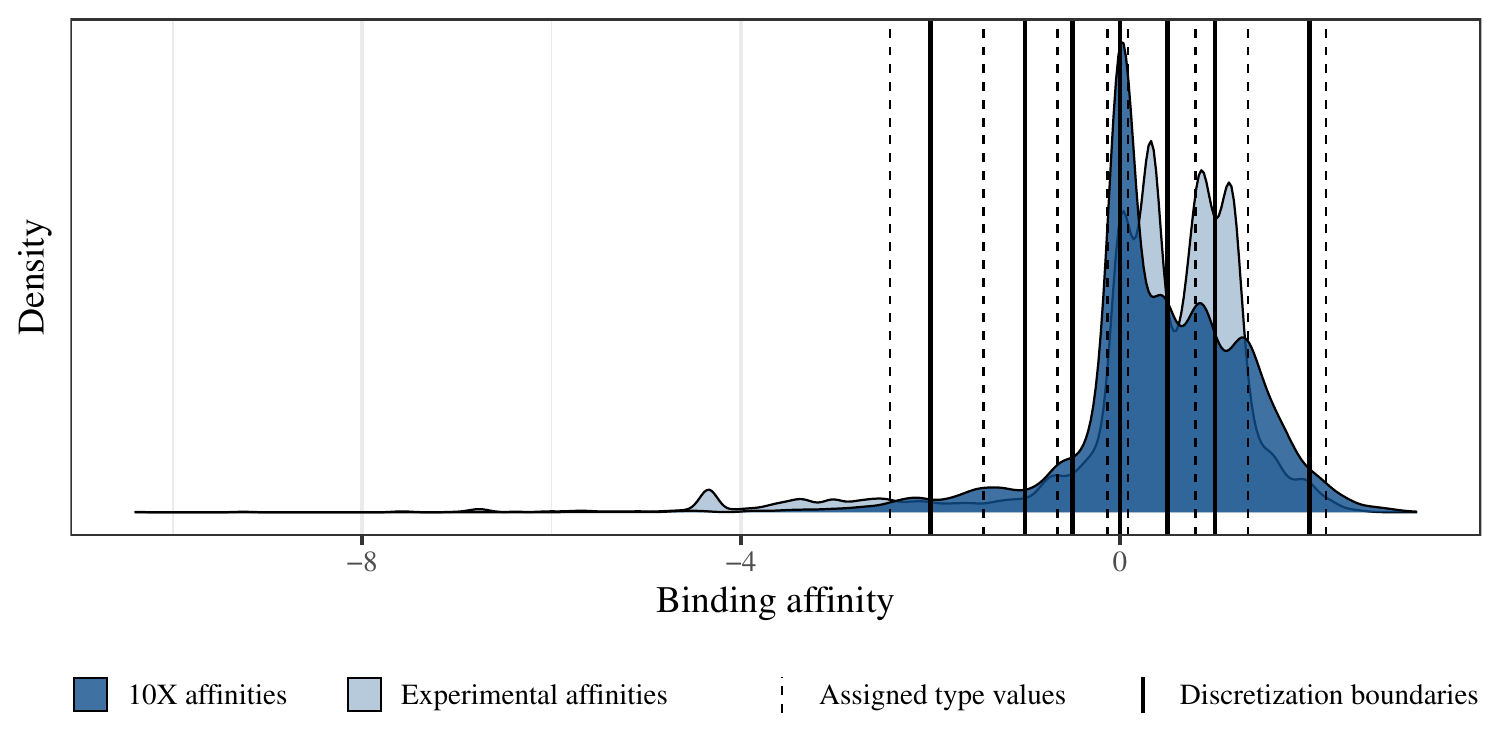}
	\caption{
		Density plots of the binding affinities seen in the 10X data and from one tree sample for each germinal center in our experimental dataset.
		The boundaries of the bins used to create the type space are marked in black.
		Type values (bin medians) are represented by dashed lines.
		Any affinities in our experimental dataset that lie outside the extreme boundaries are assigned to the extreme bins.
	}
	\label{fig:type-space}
\end{figure}

As a result, we instead use a second set of data from the 10X Genomics platform, which contains BCR sequences sampled from mice sacrificed at different points in time during the affinity maturation process \cite{dewitt2025replaying}.
The difference between this 10X data and the original data is that the former does not track to which germinal center each B cell belongs.
We use the affinity values from these sequences to determine the bin values for our discretization.
Choosing $n = 8$, we obtain $\mathcal{X} = [-2.43, -1.44, -0.66, -0.13, 0.08, 0.8, 1.35, 2.18]$.
An illustration of the type space created from this discretization, as well as its juxtaposition to the affinities observed in our experimental dataset, is shown in Fig \ref{fig:type-space}.

\subsubsection*{A priori estimation of type change rate matrix} \label{sec:tc-matrix}

We now discuss our prior knowledge of the rates of change between the types in our type space.
Previous work has yielded a hotspot-aware model of nucleotide-level mutations in B cell sequences \cite{yaari2013models, cui2016model}.
We developed a simulator that, given a B cell receptor sequence, runs a continuous-time Markov process over sequences for a fixed amount of time ($t = 200$, with a mutation rate of $1$), initialized to the given sequence and transitioning between sequences according to this context-dependent mutation model.
The simulator outputs a chain of sampled mutations that occur; no birth or death events are permitted.
The resulting mutations can then be mapped to the corresponding discretized affinity values.
Finally, we can compute the maximum likelihood rate of change between each pair of affinity values using the frequencies of each transition observed in simulation.

A rate matrix inferred this way from our experimental dataset, however, would be biased by the selection process in the germinal center, similarly to the discretization problem discussed earlier.
Instead, we opt to apply this simulator to the 5,520 sequences observed in the 10X data during the timeframe of 5 to 20 days after germinal center initialization.
The resulting rate matrix, to three decimal places, is:
\[
\boldsymbol{\Gamma} = \frac{1}{1000} \begin{pmatrix}
-4.62 & 3.35 & 0.45 & 0.27 & 0.36 & 0.09 & 0.09 & 0.0 \\
202.08 & -220.91 & 13.88 & 3.17 & 0.99 & 0.2 & 0.59 & 0.0 \\
110.3 & 150.73 & -288.49 & 22.63 & 4.61 & 0.22 & 0.0 & 0.0 \\
92.16 & 51.13 & 98.46 & -285.71 & 36.19 & 7.77 & 0.0 & 0.0 \\
86.54 & 24.62 & 65.35 & 132.85 & -332.07 & 18.27 & 4.44 & 0.0 \\
72.06 & 21.89 & 15.3 & 48.68 & 100.97 & -283.56 & 24.66 & 0.0 \\
58.28 & 20.7 & 7.92 & 13.29 & 48.06 & 99.69 & -253.06 & 5.11 \\
64.6 & 29.82 & 9.94 & 9.94 & 9.94 & 14.91 & 168.96 & -308.11
\end{pmatrix}
\]

This rate matrix is not necessarily on the same timescale as the data we analyze, as the configuration of the Markov process mutation rate here was arbitrary.
We therefore create a scaling parameter $\delta$ to be inferred, and use $\delta \boldsymbol{\Gamma}$ as the type change rate matrix in the branching process.

\subsection*{Simulation studies}

\subsubsection*{Importance of a conditional model}

Before evaluating our methodology in simulation, we first demonstrate the importance of using the conditional density in Eq (\ref{eq:tree-model}).
Consider a simple single-type branching process with constant birth rate $\lambda_\phi(\cdot) = 1.8$ and death rate $\mu = 1$, where all surviving lineages are sampled ($\rho = 1$).
Suppose we have a rejection sampling algorithm that samples trees from this process, but only accepts trees where the number of lineages alive at collection time is at least 1.
Using this sampler, we generate 100 sets of $n$ trees each.
For each tree set, we then run Metropolis-Hastings sampling within Gibbs, targeting the posterior,
\begin{equation}
p(\lambda, \mu | \mathbf{T}) \propto p(\lambda, \mu) \prod_i p_c(T_i | \lambda, \mu) ,
\end{equation}
with log-normal proposals for $\lambda$ and $\mu$ and likelihood as in Eq (\ref{eq:tree-model}), and retain posterior medians for the two rate parameters.
A $\text{log-normal}(1.5, 1)$ prior is used for $\lambda$, and a $\text{log-normal}(0, 0.5)$ prior is used for $\mu$.

The goal here is to compare the sampling distributions of the posterior median rates under this conditioned model to the corresponding distributions obtained if the unconditioned model is used, so we also run this procedure targeting the posterior for the unconditioned likelihood as in Eq (\ref{eq:tree-model-unconditioned}).
This experiment is then repeated for different sample sizes $n$ to further understand the effects of the choice of likelihood.

\begin{figure}
	\includegraphics[width=1\textwidth]{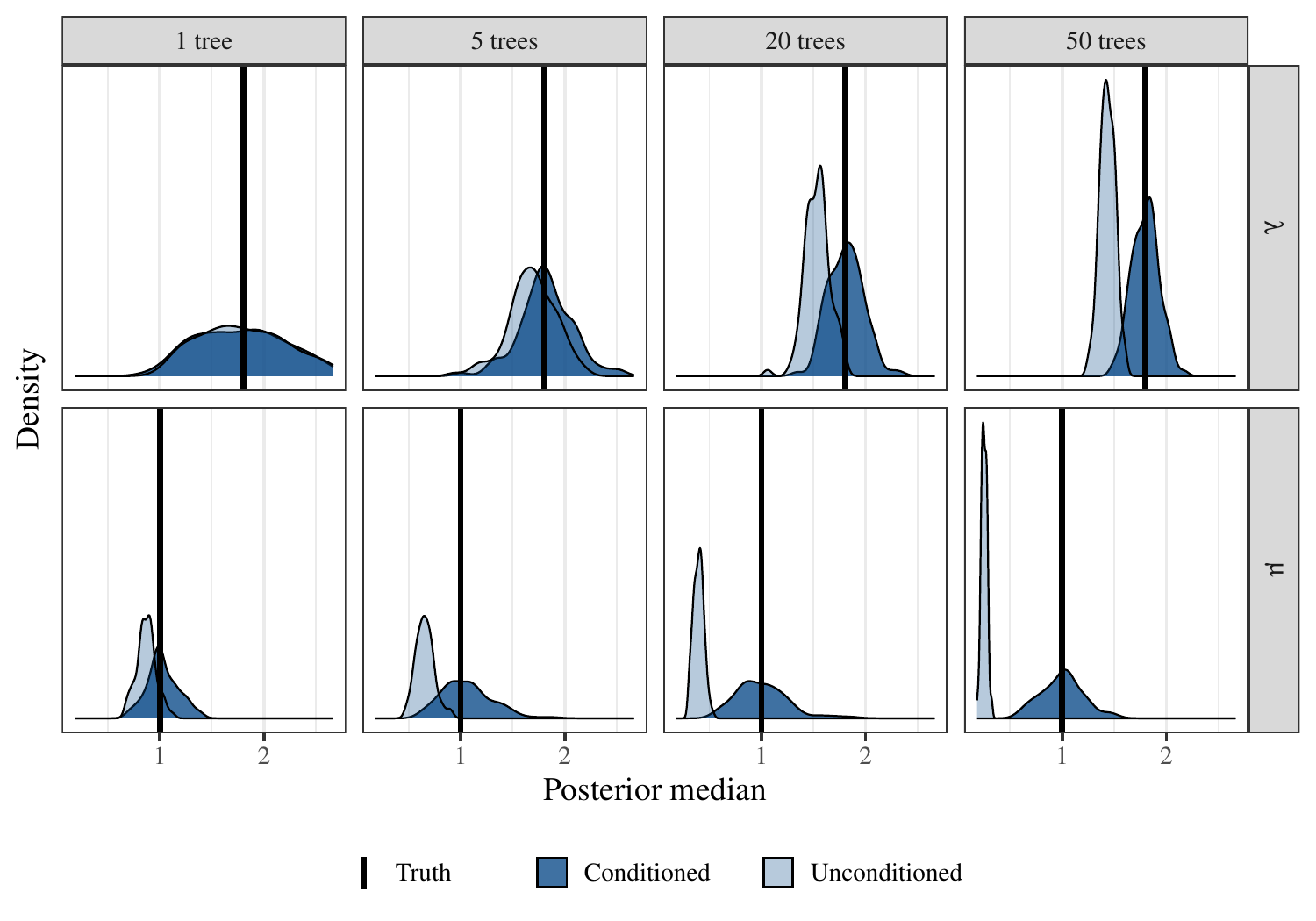}
	\caption{
		A simulation study demonstrating the importance of conditioning the tree model on the survival and observation of at least one lineage at collection time.
		Shown here are sampling distributions of posterior medians obtained from repeated runs of MCMC on different sets of trees.
		As larger tree sets are used for inference, the bias from using the unconditioned model becomes increasingly apparent, particularly for the death rate.
	}
	\label{fig:sim-study-rdt-ul}
\end{figure}

This experiment clearly demonstrates that the conditioned model mitigates a bias present when using the unconditioned model (Fig \ref{fig:sim-study-rdt-ul}).
For $n=1$, as is typical in a macroevolutionary context, we may not notice the minimal bias present by using an unconditioned tree model.
However, as the number of trees sampled increases, the bias becomes more apparent, as shown in Fig \ref{fig:sim-study-rdt-ul}.
We further note that the bias is especially extreme for the death rate parameter; the unconditioned model will tend to severely underestimate the death rate, since it will appear that cells are rarely dying.

\subsubsection*{Birth rate recovery, no model misspecification} \label{sec:sim-study-rdt-nm}

\begin{figure}
	(a) Priors

	\begin{center}
		\includegraphics[width=0.95\textwidth]{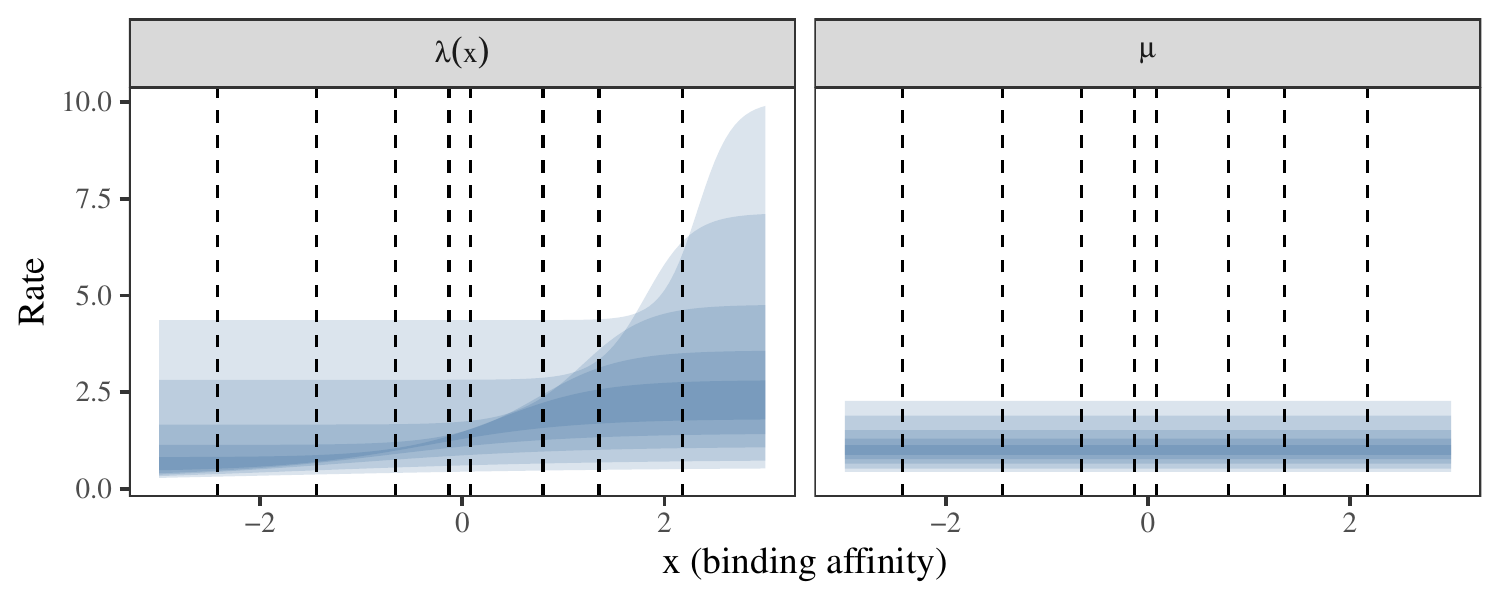}
	\end{center}

	(b) Posteriors for each simulation study

	\begin{center}
		\includegraphics[width=0.95\textwidth]{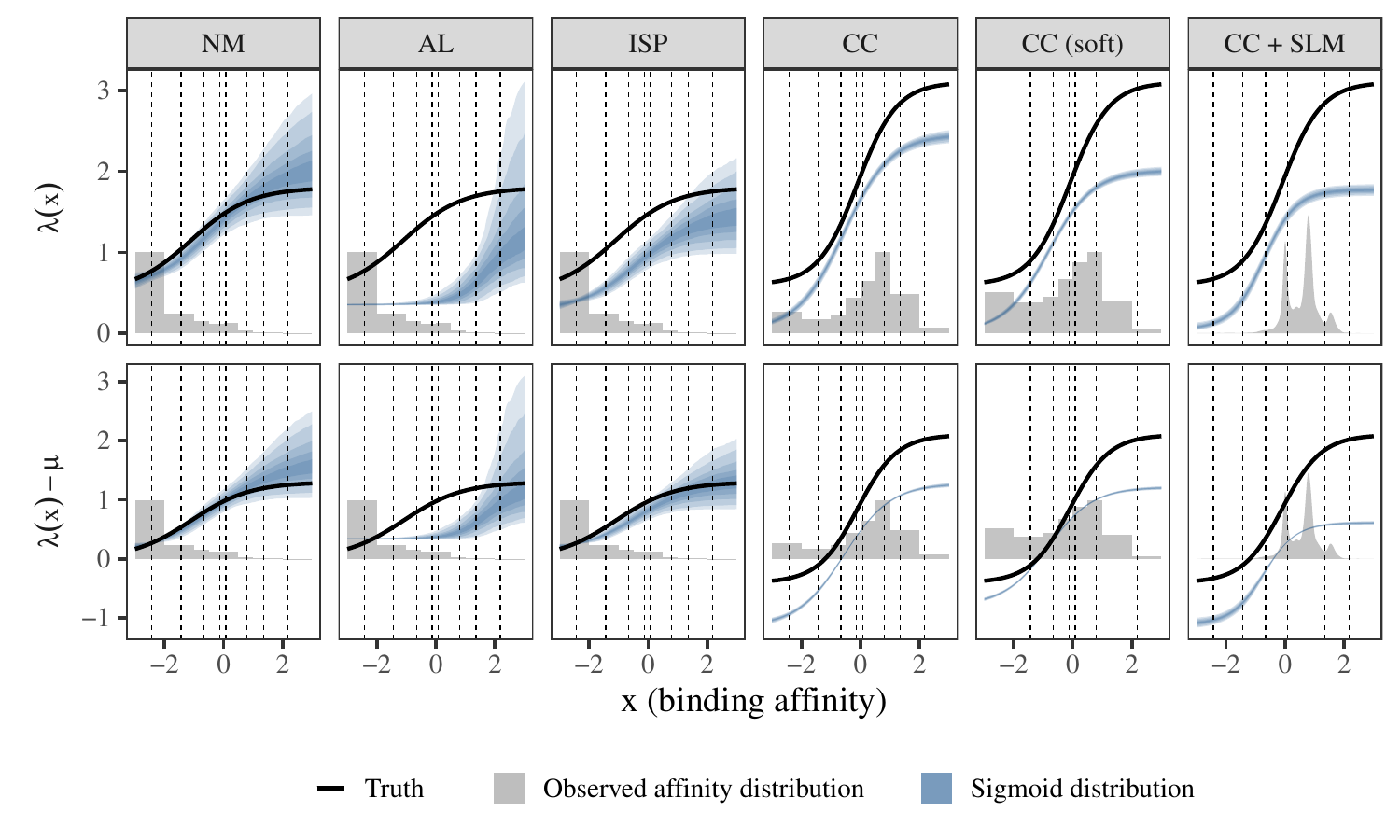}
	\end{center}
	\caption{
		\textbf{(a)} The priors over the birth rate sigmoid and death rate scalar, used for all simulation studies and data analyses.
		Shaded regions represent credible intervals (the widest being 90\%).
		The dashed lines show the values of the type space, i.e. the points at which the birth rate sigmoid is inferred.
		\textbf{(b)} The posteriors over the birth rate sigmoid for one run of each of four simulation studies, as well as the posteriors for the difference in birth rate and death rate, shown in blue.
		The abbreviations for the simulation study names are: NM, for no model misspecification; AL, for approximation of the likelihood; ISP, for incorrect sampling probability; CC, for carrying capacity; CC (soft), for soft carrying capacity; and CC+SLM, for carrying capacity with a sequence-level mutation process.
		The distributions of affinities observed in the simulated trees for each study are shown in gray.
	}
	\label{fig:sim-studies-sigmoids}
\end{figure}

We now demonstrate the performance of our inferential method in a series of simulation studies.
We begin by demonstrating our ability to recover branching process parameters when trees are simulated directly from our process.
We will then adjust the tree simulation setup to examine the effects of various model misspecifications on our inference.
The ground truth setups for all simulation studies, chosen to reflect the data observed in experimentation and our prior knowledge (see Section \textit{\nameref{sec:data}}), are documented in Appendix \textit{\nameref{sec:appendix-ss}}.
The prior distributions, chosen to enforce the sigmoidal function to be non-decreasing and non-negative, are documented in Appendix \textit{\nameref{sec:appendix-priors}} and visualized in Fig \ref{fig:sim-studies-sigmoids}(a).

The goal is to show that our parameters can be recovered from the following posterior:
\begin{equation} \label{eq:sim-study-posterior}
	p(\boldsymbol{\theta} | \mathbf{T}) \propto p(\boldsymbol{\theta}) \prod_i  p_c(T_i | \boldsymbol{\theta}) .
\end{equation}

We generated 5 sets of 58 trees each, with each tree being the result of running the branching process for 15 units of time, to mimic our experimental setup.
Trees that did not yield at least one surviving lineage to 15 units of time were discarded.
The posterior distributions of birth rate and net birth rate curves (respectively, $\lambda(x)$ and $\lambda(x) - \mu$), compared to the truth, are shown for one tree set in Fig \ref{fig:sim-studies-sigmoids}(b), ``NM'' column.
Posterior histograms and traceplots of individual parameters, as well as the sigmoid posteriors for the remaining tree sets, are shown in Appendix \textit{\nameref{sec:appendix-ss}}.
We see that the true birth rate sigmoid curve is contained within the 90\% posterior credible band for birth rate curves.

\subsubsection*{Birth rate recovery, approximation of the likelihood} \label{sec:sim-study-rdt-al}

As mentioned in Section \textit{\nameref{sec:related-work}}, Barido-Sottani et al. (2020) derive an approximate analytic solution to the multitype branching process, using an assumption that type changes do not occur in unsampled lineages.
They motivate this approximation with analysis showing sufficiently good accuracy in their scientific context, paired with large computational performance gains compared to the numerical integration necessary to compute the correct likelihood.
We implemented this method and determined that their assumption does not hold for our context.
To see this, we run the simulation study in Section \textit{\nameref{sec:sim-study-rdt-nm}} again, but during inference, we use the approximate branching process likelihood instead of the correct likelihood (which requires numerical integration).
The posterior distributions of sigmoid birth rate curves, compared to the truth, are shown for one tree set in Fig \ref{fig:sim-studies-sigmoids}(b), ``AL'' column.
Posterior histograms and traceplots of individual parameters, as well as the sigmoid posteriors for the remaining tree sets, are shown in Appendix \textit{\nameref{sec:appendix-ss}}.
We see that the posterior over birth rate curves fails to capture the true curve at many points along the affinity ($x$) axis.

\subsubsection*{Birth rate recovery, incorrect sampling probability} \label{sec:sim-study-rdt-isp}

To understand the robustness of our model to an incorrectly-known sampling probability $\rho$, we run the simulation study in Section \textit{\nameref{sec:sim-study-rdt-nm}} again, but during inference, we incorrectly set $\rho = 0.2$ instead of $\rho = 0.1$.
The posterior distributions of sigmoid birth rate curves, compared to the truth, are shown for one tree set in Fig \ref{fig:sim-studies-sigmoids}(b), ``ISP'' column.
Posterior histograms and traceplots of individual parameters, as well as the sigmoid posteriors for the remaining tree sets, are shown in Appendix \textit{\nameref{sec:appendix-ss}}.
We notice an underestimation bias of the birth rate curves, but note that the \textit{net} birth rate curve is well-recovered.

\subsubsection*{Birth rate recovery, carrying capacity} \label{sec:sim-study-cc}

Scientific consensus is that the germinal center has a carrying capacity for B cells.
In the late stages of affinity maturation, the germinal center reaches a maximum cell population size and begins to inhibit the cells' proliferation, potentially regardless of their antibodies' binding affinity, to adjust evolutionary dynamics and control this population size.
To understand the robustness of our model to experimental data sampled in such a phenomenon, we run the simulation study in Section \textit{\nameref{sec:sim-study-rdt-nm}} again, but with a carrying capacity implemented.
When the number of living cells in the tree reaches 1000, a lineage is chosen uniformly randomly to die.
The posterior distributions of sigmoid birth rate curves, compared to the truth, are shown for one tree set in Fig \ref{fig:sim-studies-sigmoids}(b), ``CC'' column.
Posterior histograms and traceplots of individual parameters, as well as the sigmoid posteriors for the remaining tree sets, are shown in Appendix \textit{\nameref{sec:appendix-ss}}.
We notice an underestimation bias of both the birth rate and net birth rate curves; the branching process assumed for inference requires a lower overall birth rate to attempt to explain the observed distribution of tree sizes.

Note that the tighter posteriors in the carrying capacity studies can be attributed to the larger tree sizes used for inference in these studies.
It is considerably easier to reliably simulate large trees in the presence of a capacity, which prevents tree size explosion.

\subsubsection*{Birth rate recovery, soft carrying capacity} \label{sec:sim-study-cc-soft}

To explore the effects of different implementations of a carrying capacity, we modified and repeated the simulation study in Section \textit{\nameref{sec:sim-study-cc}}.
When the population reaches capacity in this study, rather than randomly selecting a lineage to die, we logistically modulate the birth rate such that the process is critical.
The posterior distributions of sigmoid birth rate curves, compared to the truth, are shown for one tree set in Fig \ref{fig:sim-studies-sigmoids}(b), ``CC (soft)'' column.
Posterior histograms and traceplots of individual parameters, as well as the sigmoid posteriors for the remaining tree sets, are shown in Appendix \textit{\nameref{sec:appendix-ss}}.
We notice a similar underestimation bias as in Section \textit{\nameref{sec:sim-study-cc}}, but with an additional effect of the range of the sigmoid being smaller compared to the sigmoid inferred under the hard carrying capacity model.

\subsubsection*{Birth rate recovery, carrying capacity with sequence-level mutation process} \label{sec:sim-study-cc-slm}

In the germinal center, B cell mutations occur on the sequence level; in our branching process, we assume type changes are Markovian with respect to the binding affinity, a function of the genetic sequence.
Our model thus assumes an oversimplification of the B cell mutation process that needs to be examined.
We run the simulation study in Section \textit{\nameref{sec:sim-study-cc}}, but the type space and type change rate matrix are replaced by the sequence-level mutation process described in Section \textit{\nameref{sec:tc-matrix}}.
The posterior distributions of sigmoid birth rate curves, compared to the truth, are shown for one tree set in Fig \ref{fig:sim-studies-sigmoids}(b), ``CC+SLM'' column.
Posterior histograms and traceplots of individual parameters, as well as the sigmoid posteriors for the remaining tree sets, are shown in Appendix \textit{\nameref{sec:appendix-ss}}.
We notice a similar underestimation bias and reduced sigmoid range as in Section \textit{\nameref{sec:sim-study-cc-soft}}.

\subsection*{Analysis of experimental data} \label{sec:data}

\subsubsection*{Data description} \label{sec:data-description}

In experimentation, 18 genetically identical mice genetically unable to generate their own GCs received adoptive transfers of GC-competent monoclonal B cells specific for chIgY and were then immunized with this antigen to trigger GC formation.
We observed the results of 52 parallel replications of affinity maturation---one for each germinal center sampled across the mice.
B cells were sampled from the mice after 15 days, and the \textit{Igh}2.1 and \textit{Igk}2.1 alleles were sequenced with the Illumina MiSeq platform as described in DeWitt, Vora, Taraki et al. (2025) \cite{dewitt2025replaying}.
Sequences were aligned and concatenated, with the final 3,758 alignments having length 657 and no insertions or deletions.

As discussed in Section \textit{\nameref{sec:problem}}, we need to be able to estimate a phylogenetic tree with type annotations for each germinal center.
To do so, we use the software package BEAST to sample 5,000 (after chain thinning) trees for each germinal center, from the posterior formed with the HKY substitution model and generalized Bayesian Skyline tree prior \cite{suchard2018bayesian, hasegawa1985dating, drummond2005bayesian}.
To obtain type change annotations, we use a feature implemented in BEAST for sampling ancestral states, as well as mutations which generate these ancestral states \cite{nielsen2002mapping, hobolth2009simulation}.
Since we currently do not have the machinery to jointly infer rate parameters and phylogenies, we uniformly randomly select one tree from the resulting MCMC chain for each germinal center to use for inference.

Notably, we enforced each sampled tree to have the naive sequence at the root, since we controlled for this in experimentation.
To do so, we introduced the naive sequence as a sampled leaf when configuring BEAST, setting the sampling time to a small positive $\epsilon \approx 0$.
We then ``inverted'' the small $\epsilon$-length branch in each sampled tree to have the naive sequence at the root.
An example of one sampled tree from two different germinal centers is shown in Fig \ref{fig:example-beast-trees}.

\subsubsection*{Parameter inference}

We fix $\boldsymbol{\Gamma}$ and $\rho_i$ as discussed in Section \textit{\nameref{sec:adapting}} and conduct inference of $\boldsymbol{\theta}$.
Priors used are documented in Appendix \textit{\nameref{sec:appendix-priors}}.
The posterior distribution of sigmoid birth rate curves is shown in Fig \ref{fig:data-analysis}.
Posterior histograms and traceplots of individual parameters are shown in Appendix \textit{\nameref{sec:appendix-da}}.

\begin{figure}
	\includegraphics[width=1\textwidth]{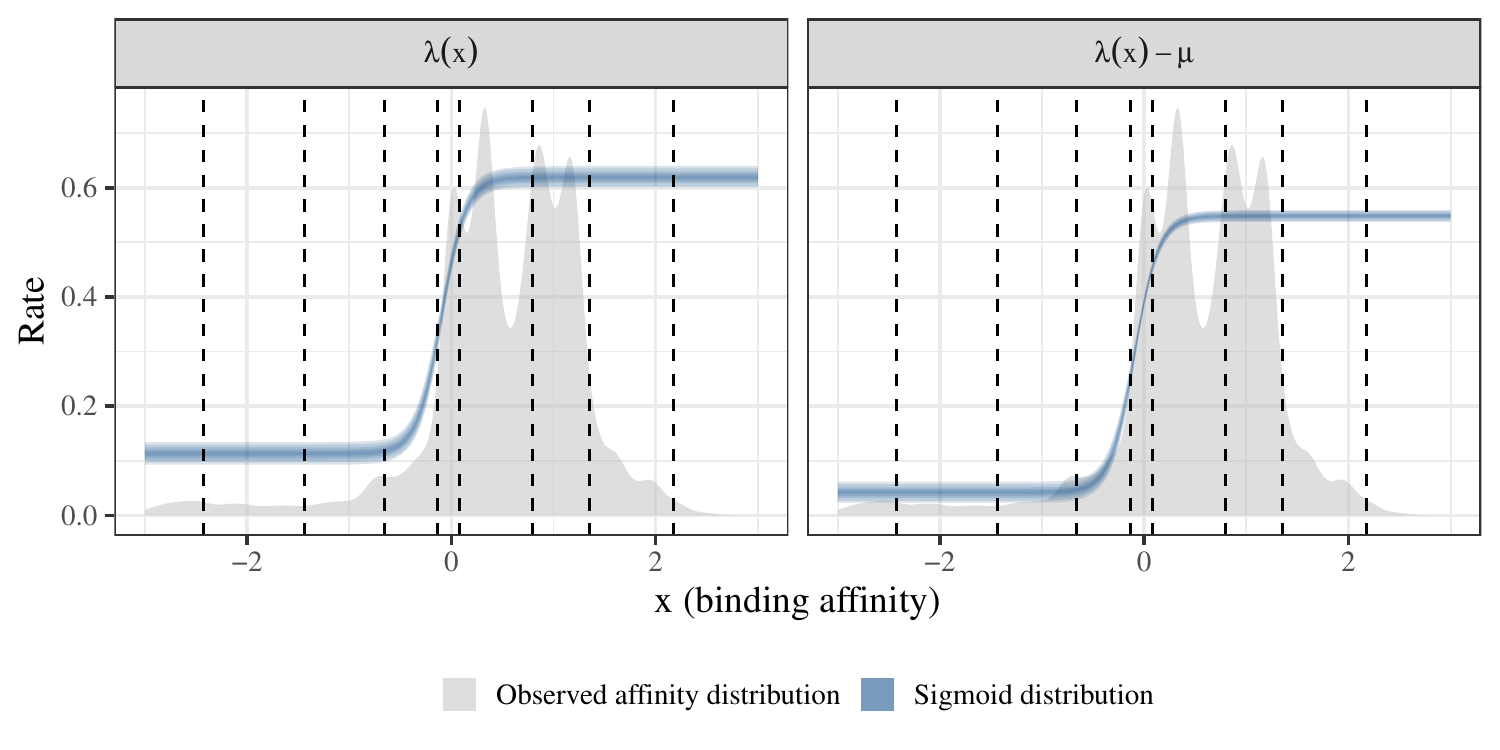}

	\caption{
		A comparison of the prior and posterior over sigmoid curves for the experimental data analysis, shown in blue.
		Shaded regions represent credible intervals (the widest being 90\%).
		The dashed lines show the values of the type space, i.e. the points at which the birth rate sigmoid is inferred.
		The distribution of affinities observed in the BEAST trees is shown in gray.
	}
	\label{fig:data-analysis}
\end{figure}

We observe a sigmoid with a range roughly spanning the interval (0.1, 0.6), and with a steep transition near the zero-valued $-\Delta \log_{10} K_D$ region.
In light of the observed effects of carrying capacity in our simulation studies, we suspect that these birth rate curves are underestimated.
Nonetheless, the order of magnitude of the inferred birth rates corresponding to different BCR binding affinities is likely reasonable.

Recall from Section \textit{\nameref{sec:data-description}} that we use one tree sample per germinal center to conduct inference.
This behavior ignores the phylogenetic uncertainty captured in the chain of tree samples from BEAST.
To investigate the effect of ignoring this uncertainty, we ran additional analyses, using different tree samples for each germinal center.
The results are also shown in Appendix \textit{\nameref{sec:appendix-da}}; we found the effect to be relatively minor.
Some variation was observed in the posteriors for the birth rate curve, but the overall range of the curve remained consistent.

\subsubsection*{Posterior predictive diagnostics} \label{sec:post-pred-check}

To examine how well our model fits the data, we perform posterior predictive diagnostics.
Each posterior sample is used to generate a replicate set of germinal center data, by simulating 52 trees and collecting the sequences at the leaves.
For each replicate, we produce 8 data summary statistics---the proportion of leaves with each of the 8 discretized binding affinity values, aggregated over all the trees.
The distribution of each summary, compared to the observed proportion of each affinity value in the experimental data, is visualized in Appendix \textit{\nameref{sec:appendix-da}}.
We note how the frequency of the fourth-largest affinity is overestimated, while those of the second and third-largest are underestimated.
This indicates that the sigmoid inferred under our model may be too steep.
We suspect this could be a consequence of not modeling the germinal's carrying capacity.

\section*{Discussion}

In this paper, we developed a multitype branching process with the flexibility to model a functional relationship between birth rates and lineage attributes.
We then applied this model to a novel experiment studying antibody affinity maturation to illuminate how antibody binding affinity translates into B cell fitness in the germinal center.
We reiterate that this relationship cannot be measured directly and poses a long-standing problem in the field of immunology \cite{KURAOKA2016542, BATISTA1998751}.

In simulation studies, we found that a sigmoidal mapping of affinity to B cell birth rate could be recovered if the data-generating process is indeed our branching process, but showed the biases that can occur under a more biologically-accurate data-generating process with a carrying capacity.
We also found that the presence of replicate evolutionary processes contributes to an underestimation bias of the birth and death rates if one does not condition on non-extinction.
Finally, we observed that the approximate likelihood of Barido-Sottani et al. (2020) did not perform well in our context.
This may be a result of the large number of lineages that go unobserved, whether due to the B cells being eliminated or merely going unsampled.

In data analysis, we inferred a sigmoidal mapping that indicates a factor of 6 difference in birth rate for high-affinity vs low-affinity B cells.
Seeing how the carrying capacity simulation studies indicate a potential bias of vertical compression for the sigmoid curve, we conjecture that the true difference is even larger than a factor of 6.
While this analysis has limitations, which we address below, to our knowledge this is the first phylogenetic-based attempt to infer the affinity-fitness relationship in the germinal center from BCR sequence data.

There are a few limitations of our method that, if overcome, could provide for a more biologically faithful analysis of this germinal center data.
First, we are not jointly inferring the birth rate mapping with the phylogenetic trees, and use only one tree per germinal center for a given analysis.
This could underestimate uncertainty around our branching process parameters (also see Appendix \textit{\nameref{sec:appendix-da}}).
We could conduct joint inference by building a hierarchical model, where our branching process density serves as the prior over the phylogenetic tree.
In this manner, both the tree and birth rate mapping are inferred directly from the observed genetic sequences.
The implementation of such a hierarchical model can be computationally challenging, though, as the MCMC algorithm would need to operate over a state space involving trees.
A potential alternative to this solution could be a sampling importance resampling procedure with our existing tree samples, although this would require that the prior density which generated our samples is sufficiently similar to our branching process density \cite{gordon1993novel, liang2007hierarchical}.

Second, as discussed in Section \textit{\nameref{sec:sim-study-cc}} and Section \textit{\nameref{sec:post-pred-check}}, there is a consequential model misspecification when not accounting for the carrying capacity of the germinal center.
DeWitt, Vora, Taraki et al. (2025) used a non-phylogenetic traveling wave model that accounts for carrying capacity to infer the affinity-fitness relationship \cite{dewitt2025replaying}. 
However, this model was fit to bulk, not single-cell, data collected at multiple time points. 
The work of Ralph et al. (2025) provides a framework more comparable to ours \cite{ralph2025inference}.
These authors use deep learning and simulation-based inference to obtain the affinity-fitness relationship under a branching process model with a carrying capacity constraint, enforced by a birth rate modulated by the number of germinal center cells.
This is the same model we use in our simulation study described in Section \textit{\nameref{sec:sim-study-cc-soft}}.
Ralph et al. (2025) discuss similarities and differences between our birth rate and their effective/modulated birth rate in their Discussion section. 
In short, the main difference is that our sigmoidal inference shows a sharp increase in birth rate as the BCR binding affinity increases, while Ralph et al. (2025) infer a more gradual increase in effective birth rate.
They suggest that our sharp increase could be a result of not modeling carrying capacity, which is plausible.

One experimental way our branching process framework could avoid the effects of this misspecification is by collecting B cell sequence data at an earlier point in time during the affinity maturation process, before the germinal center achieves its capacity.
A more model-based solution, though, could be to allow for time-varying rate parameters in the branching process.
Stadler (2011) describes a branching process in which the birth and death rates are piecewise-constant over a set of time intervals \cite{stadler2011mammalian}.
Magee et al. (2020) brings this work into a Bayesian context, exploring both Gaussian and horseshoe Markov random field priors for the rates \cite{magee2020locally}.
One could lower the y-scale parameter ($\phi_1$) of the birth rate sigmoid in this manner, at a time at which carrying capacity has kicked in.
Alternatively, recent work by DeWitt et al. (2024) discuss how the branching process with a carrying capacity can have a tractable likelihood if a mean-field approach to modeling lineage interactions is taken \cite{dewitt2024meanfieldinteractingmultitypebirthdeath}.

Finally, future work could improve the manner in which we specify the affinity-fitness relationship.
In this work, we specifically assumed a parametric sigmoidal form for the birth rate function.
This form may not hold in practice.
Instead, a Bayesian non-parametric approach could provide the flexibility necessary to avoid potential misspecification of the function; Gaussian processes or adaptive spline methods could prove useful for this task.
Moreover, to relax the assumption that the same functional form governs the affinity-fitness relationship across mice, or even across individual germinal centers, Bayesian hierarchical techniques could be used to capture this potential heterogeneity.

\section*{Acknowledgments}
This material is based upon work supported by the National Science Foundation Graduate Research Fellowship Program under Grant No DGE-1839285.
Any opinions, findings, and conclusions or recommendations expressed in this material are those of the authors and do not necessarily reflect the views of the National Science Foundation.

\bibliography{../../references}

\pagebreak

\appendix

{\Large
\textbf\newline{\bf Supplementary Materials}
}
\newline

\section*{Additional simulation study details} \label{sec:appendix-ss}

\subsection*{No model misspecification}

The following parameter values and figures correspond to the simulation study described in Section \textit{\nameref{sec:sim-study-rdt-nm}}.
The parameter values configure the data generating process used to simulate each tree for the 5 tree sets used in the study.
The first figure demonstrates the performance of our method by visualizing the posterior over sigmoid curves inferred for each tree set.
The second figure decomposes this posterior visualization into posterior histograms for each sigmoid parameter individually, as well as for the other branching process parameters.
Finally, the third figure shows traceplots for each of these parameters.
The subsequent sections of this appendix for each other simulation study are structured in the same manner.

\subsubsection*{Ground truth parameter values}

\begin{itemize}
	\item $\phi = [1.3, 1, -1.1, 0.5]$
    \item $\mu = 0.5$
    \item $\delta = 20$
    \item $\rho = 0.1$
\end{itemize}

The type space and type change rate matrix are taken from Sections \textit{\nameref{sec:discretization}} and \textit{\nameref{sec:tc-matrix}}, respectively.

\subsubsection*{Additional figures}

\begin{figure}[H]
	\centering
	\includegraphics[width=1\textwidth]{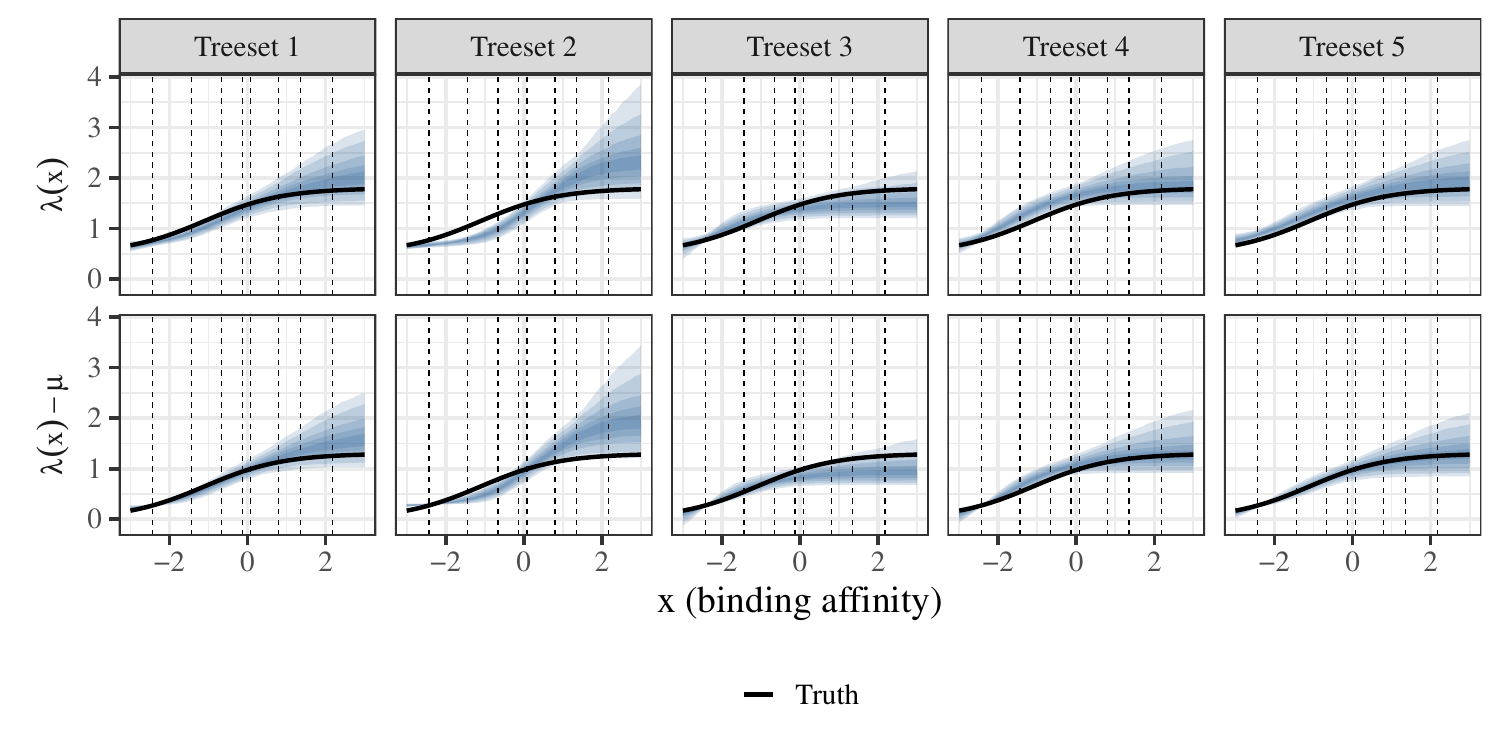}
	\caption{
		\textbf{(No model misspecification)}
		Posteriors of the birth rate sigmoid and difference in birth rate and death rate, for each tree set.
		Shaded regions represent credible intervals (the widest being 90\%).
		The dashed lines show the values of the type space, i.e. the points at which the birth rate sigmoid is inferred.
	}
\end{figure}

\begin{figure}[H]
	\centering \small
	\includegraphics[width=1\textwidth]{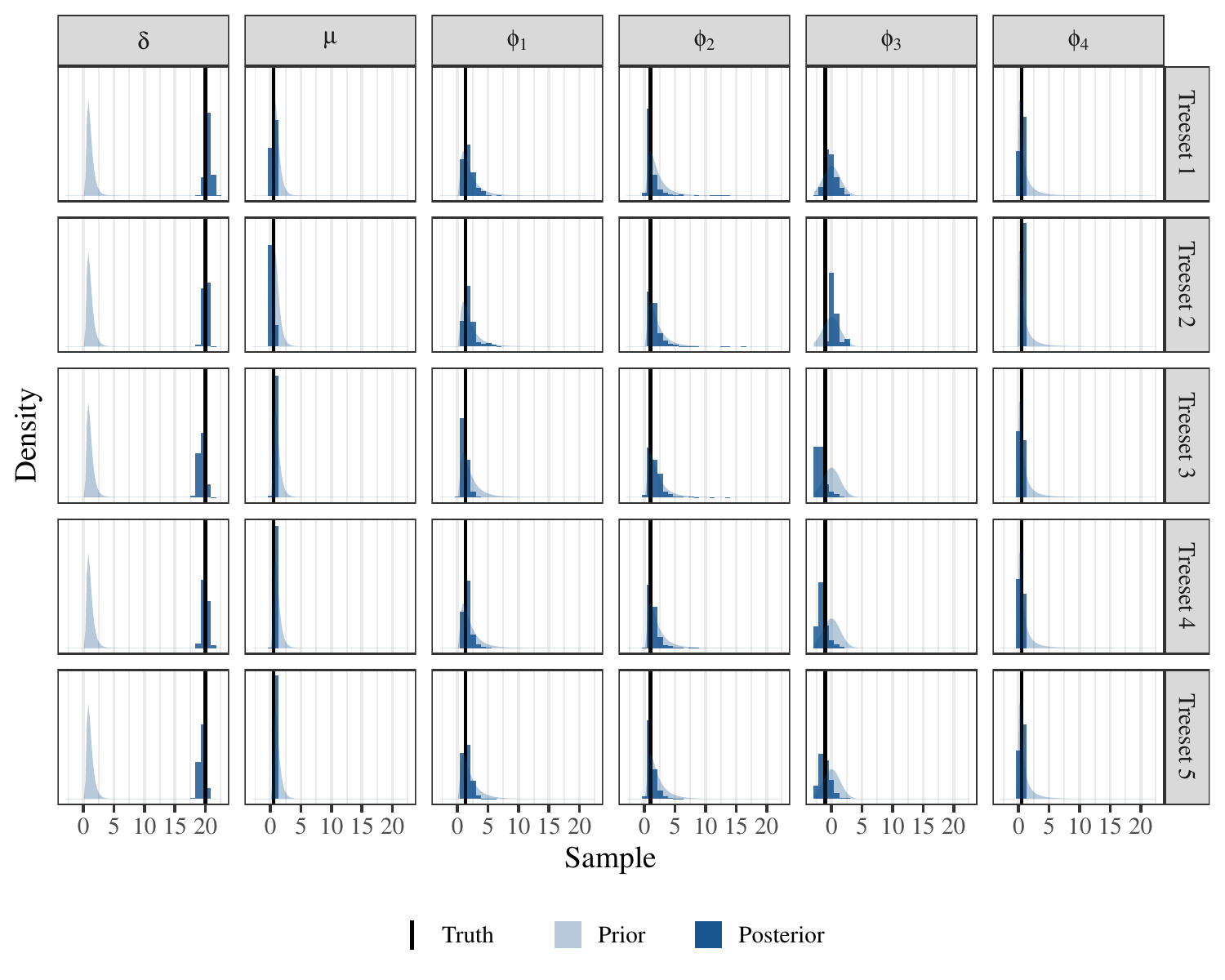}
	\caption{
		\textbf{(No model misspecification)}
		Posterior histograms and prior density curves of each parameter, for each tree set.
	}
\end{figure}

\begin{figure}[H]
	\centering \small
	\includegraphics[width=1\textwidth]{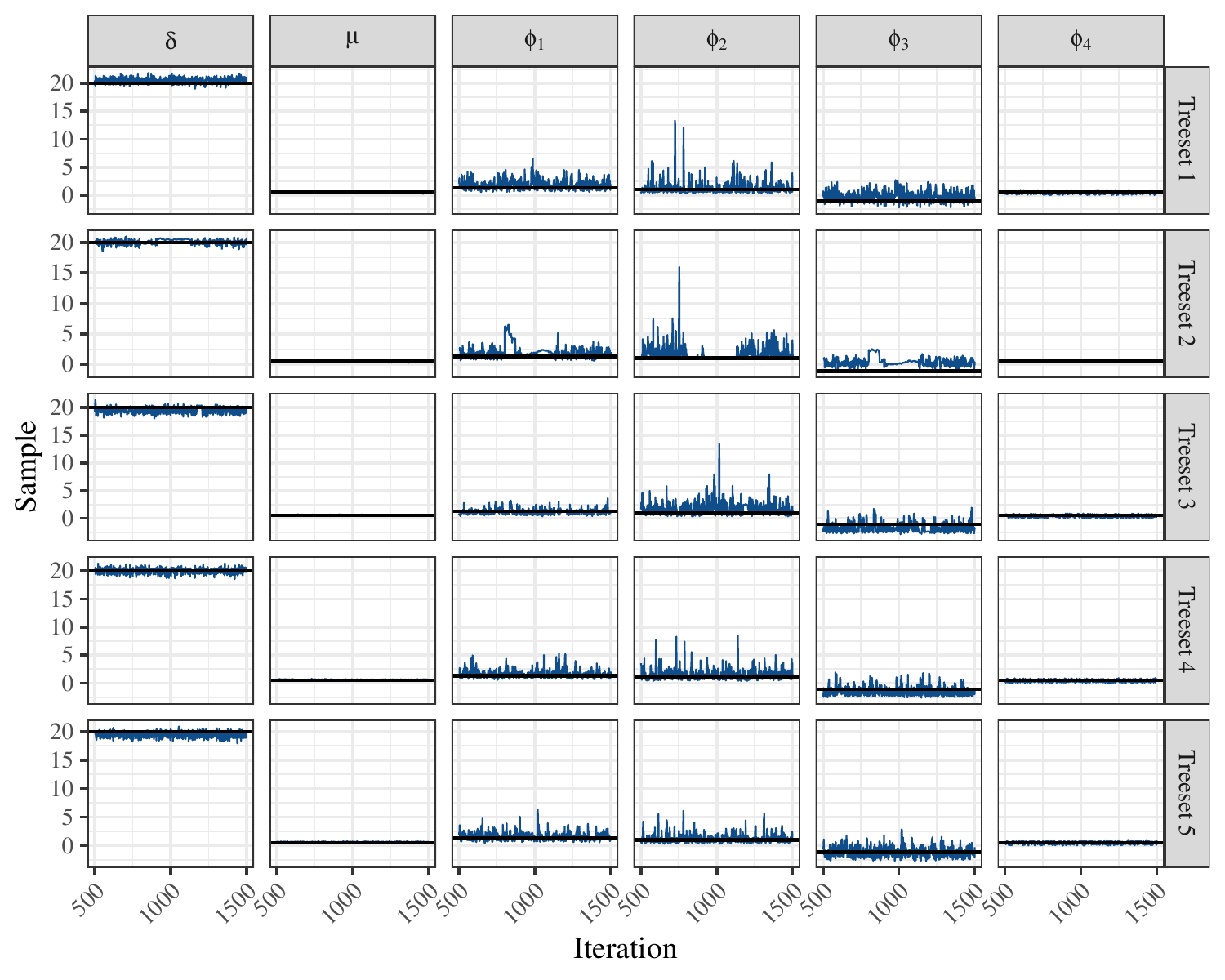}
	\caption{
		\textbf{(No model misspecification)}
		Posterior traceplots of each parameter, for each tree set.
	}
\end{figure}

\pagebreak

\subsection*{Approximation of the likelihood}

The following parameter values and figures correspond to the simulation study described in Section \textit{\nameref{sec:sim-study-rdt-al}}.

\subsubsection*{Ground truth parameter values}

\begin{itemize}
	\item $\phi = [1.3, 1, -1.1, 0.5]$
    \item $\mu = 0.5$
    \item $\delta = 20$
    \item $\rho = 0.1$
\end{itemize}

The type space and type change rate matrix are taken from Sections \textit{\nameref{sec:discretization}} and \textit{\nameref{sec:tc-matrix}}, respectively.

\subsubsection*{Additional figures}

\begin{figure}[H]
	\centering
	\includegraphics[width=1\textwidth]{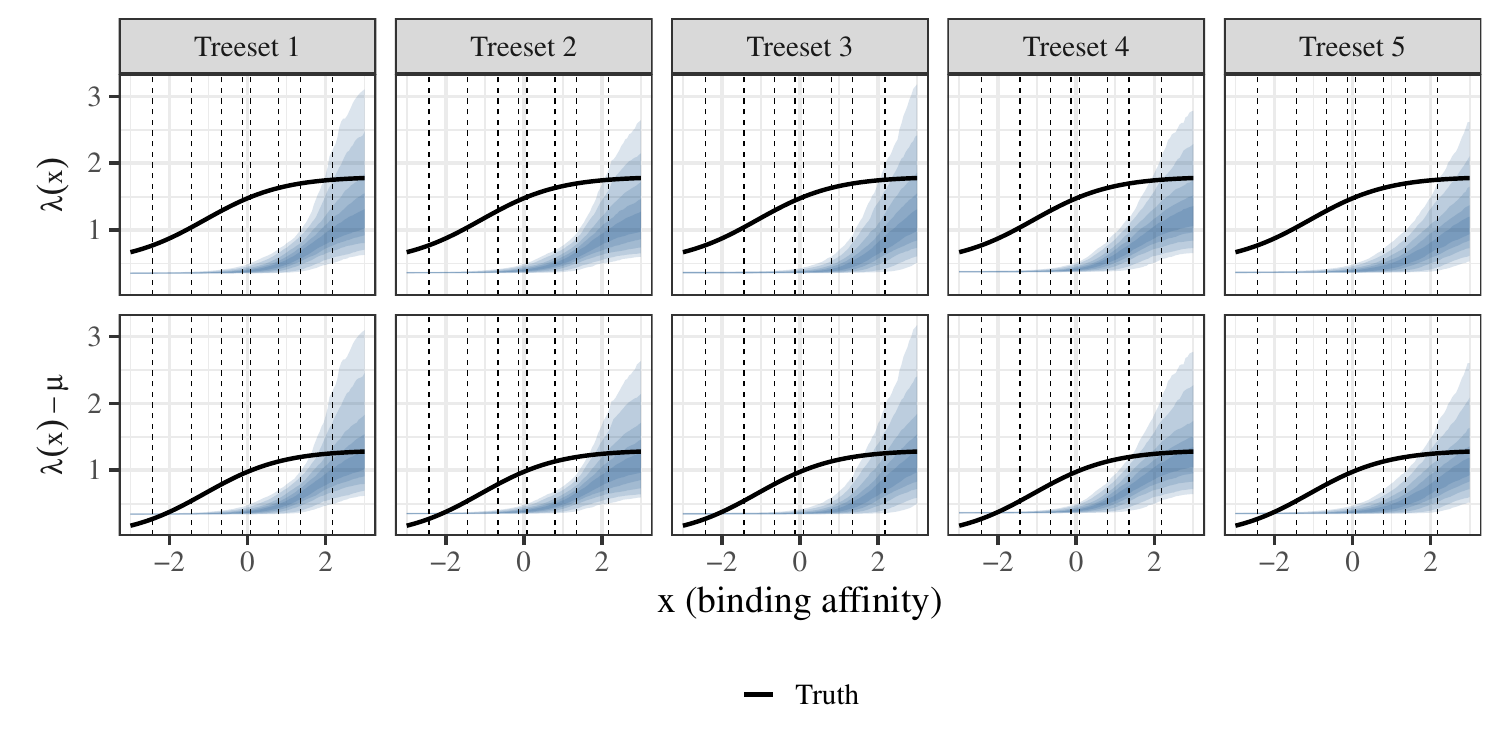}
	\caption{
		\textbf{(Approximation of the likelihood)}
		Posteriors of the birth rate sigmoid and difference in birth rate and death rate, for each tree set.
		Shaded regions represent credible intervals (the widest being 90\%).
		The dashed lines show the values of the type space, i.e. the points at which the birth rate sigmoid is inferred.
	}
\end{figure}

\begin{figure}[H]
	\centering \small
	\includegraphics[width=1\textwidth]{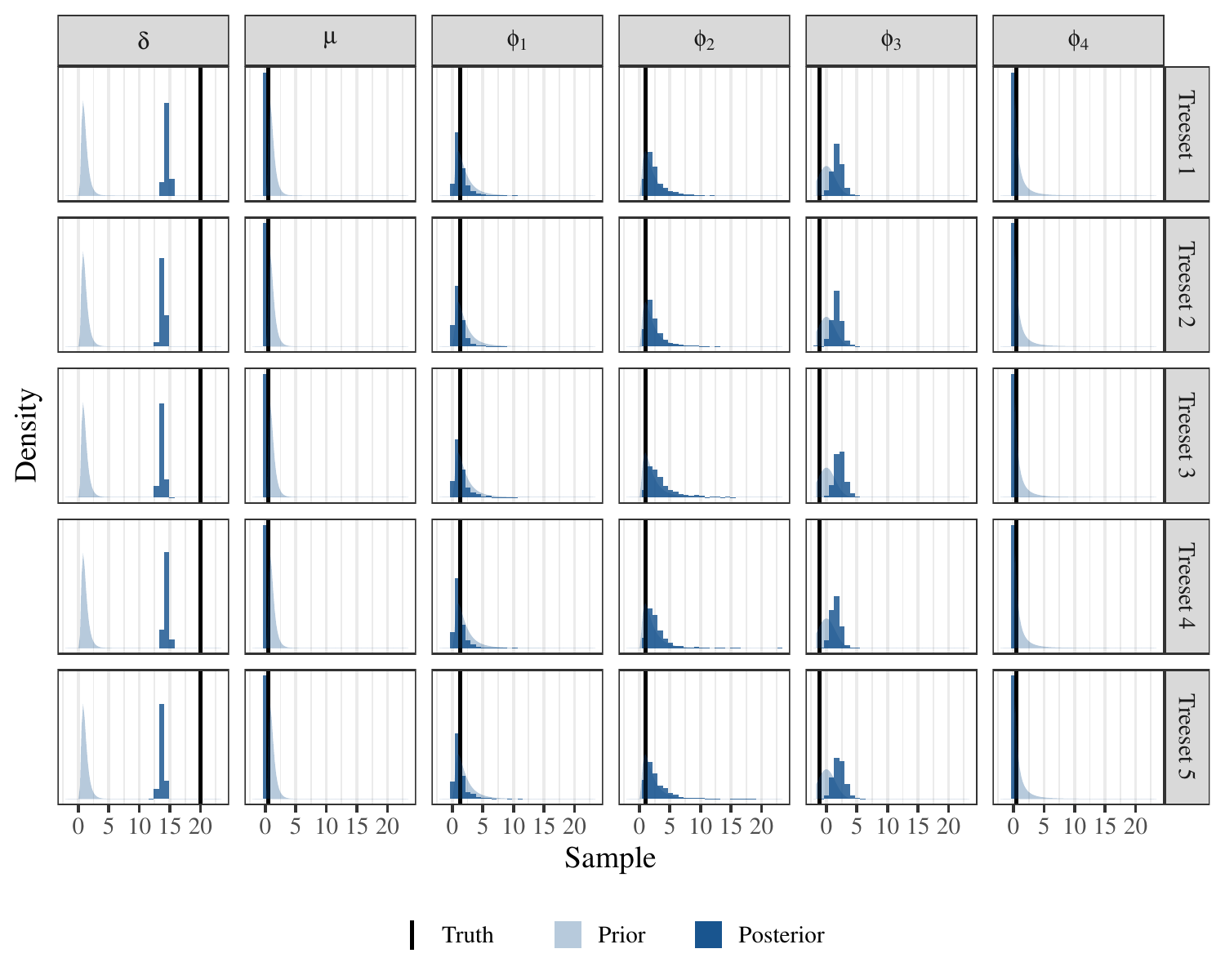}
	\caption{
		\textbf{(Approximation of the likelihood)}
		Posterior histograms and prior density curves of each parameter, for each tree set.
	}
\end{figure}

\begin{figure}[H]
	\centering \small
	\includegraphics[width=1\textwidth]{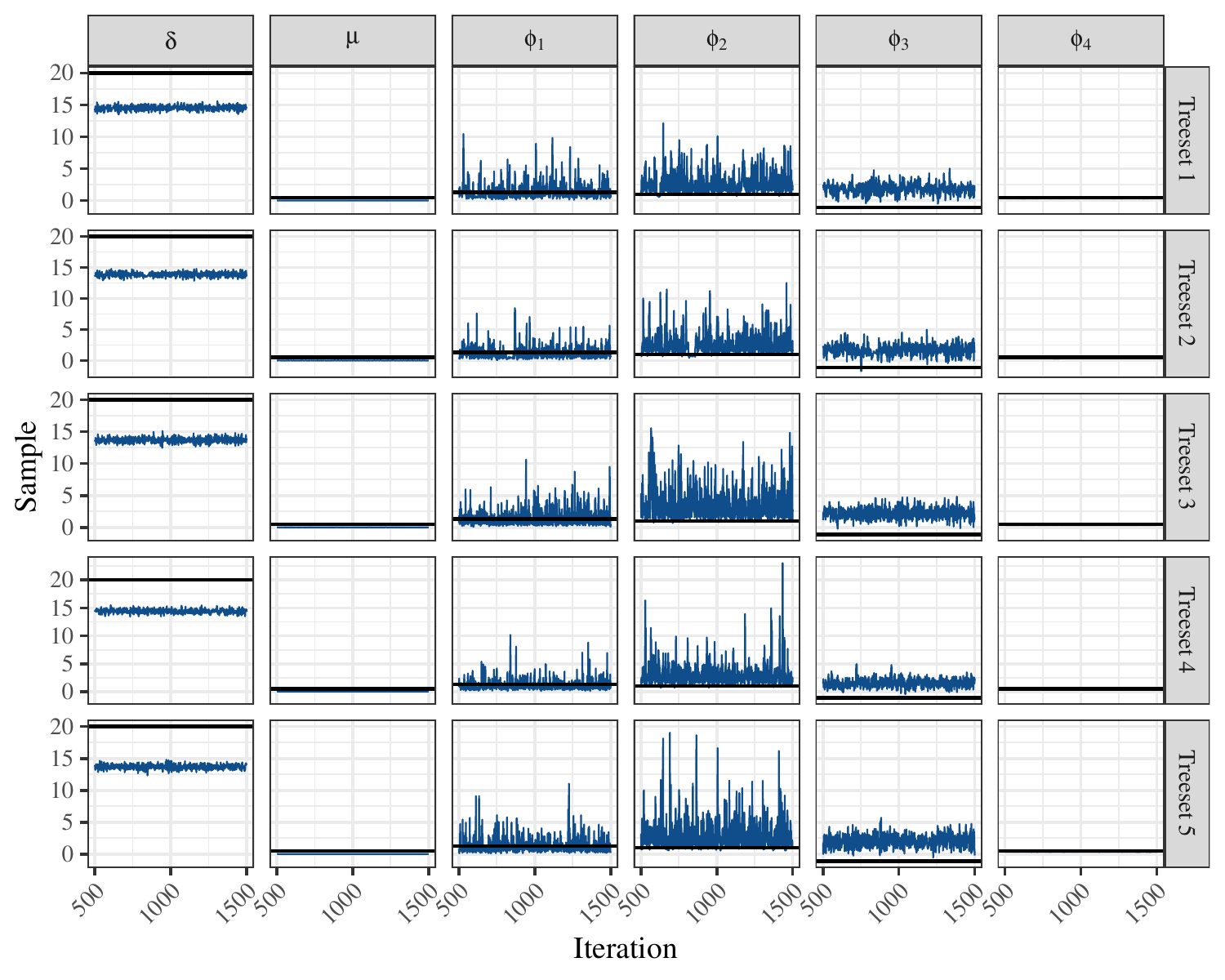}
	\caption{
		\textbf{(Approximation of the likelihood)}
		Posterior traceplots of each parameter, for each tree set.
	}
\end{figure}

\pagebreak

\subsection*{Incorrect sampling probability}

The following parameter values and figures correspond to the simulation study described in Section \textit{\nameref{sec:sim-study-rdt-isp}}.

\subsubsection*{Ground truth parameter values}

\begin{itemize}
	\item $\phi = [1.3, 1, -1.1, 0.5]$
    \item $\mu = 0.5$
    \item $\delta = 20$
    \item $\rho = 0.1$
\end{itemize}

The type space and type change rate matrix are taken from Sections \textit{\nameref{sec:discretization}} and \textit{\nameref{sec:tc-matrix}}, respectively.

During inference, $\rho$ is set to be $0.2$.

\subsubsection*{Additional figures}

\begin{figure}[H]
	\centering
	\includegraphics[width=1\textwidth]{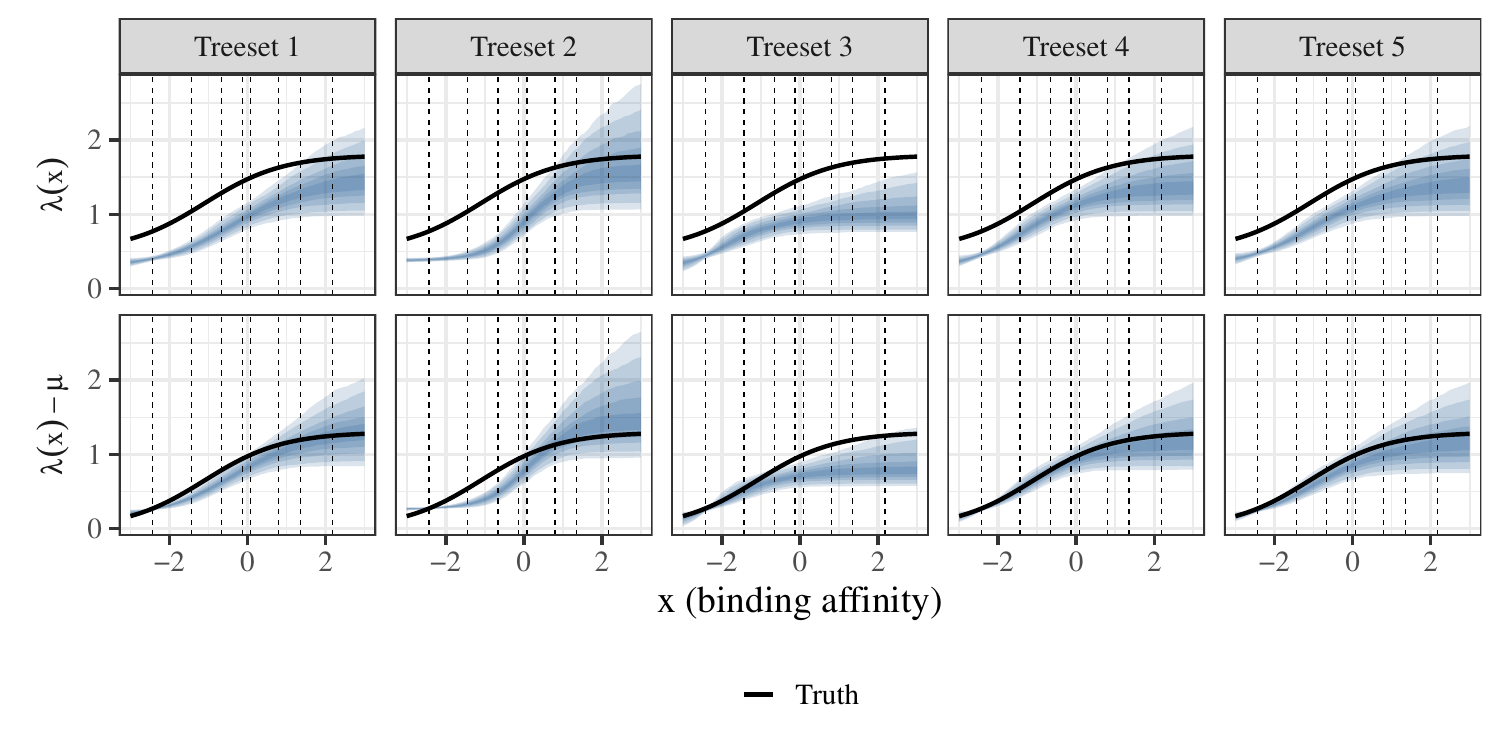}
	\caption{
		\textbf{(Incorrect sampling probability)}
		Posteriors of the birth rate sigmoid and difference in birth rate and death rate, for each tree set.
		Shaded regions represent credible intervals (the widest being 90\%).
		The dashed lines show the values of the type space, i.e. the points at which the birth rate sigmoid is inferred.
	}
\end{figure}

\begin{figure}[H]
	\centering \small
	\includegraphics[width=1\textwidth]{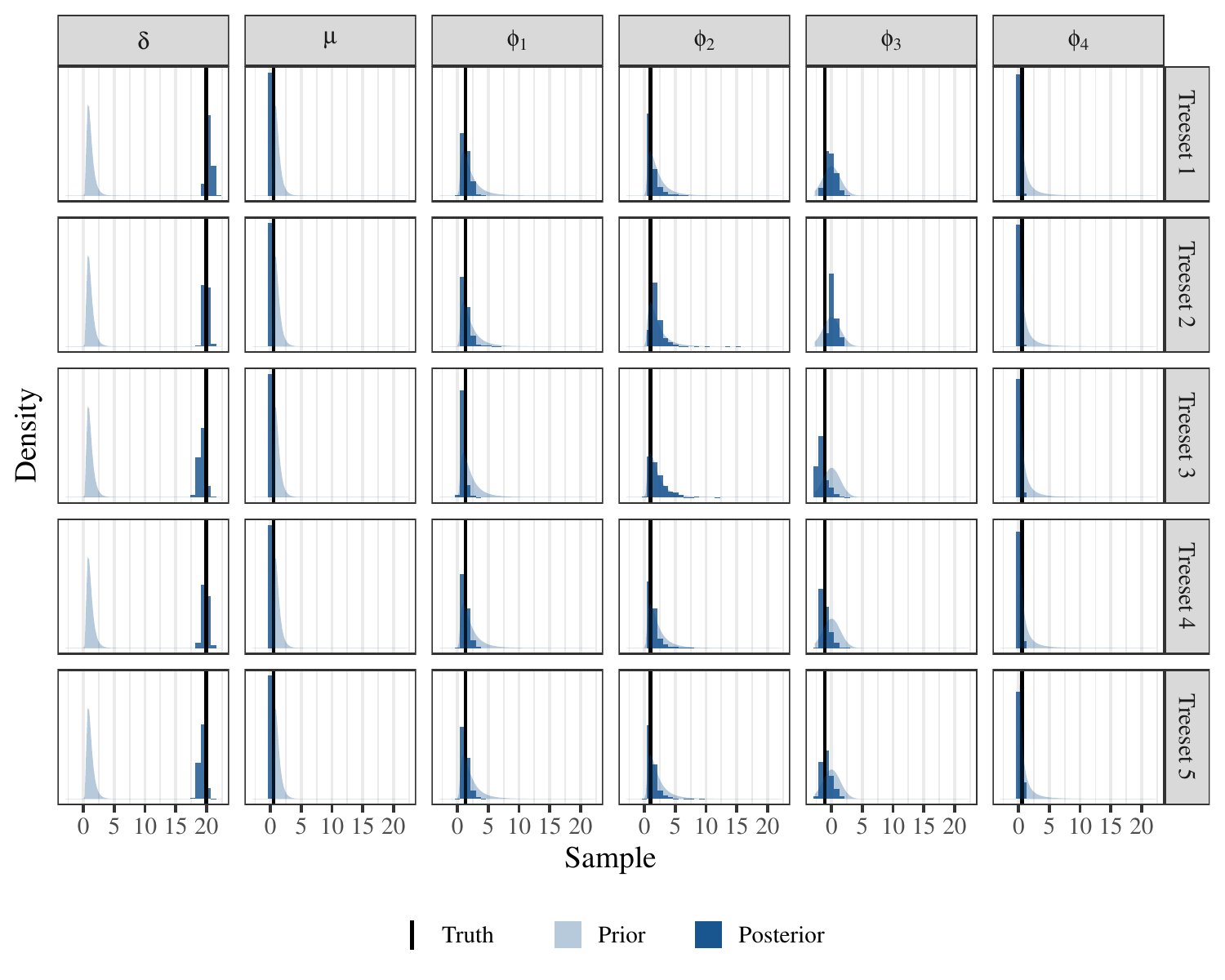}
	\caption{
		\textbf{(Incorrect sampling probability)}
		Posterior histograms and prior density curves of each parameter, for each tree set.
	}
\end{figure}

\begin{figure}[H]
	\centering \small
	\includegraphics[width=1\textwidth]{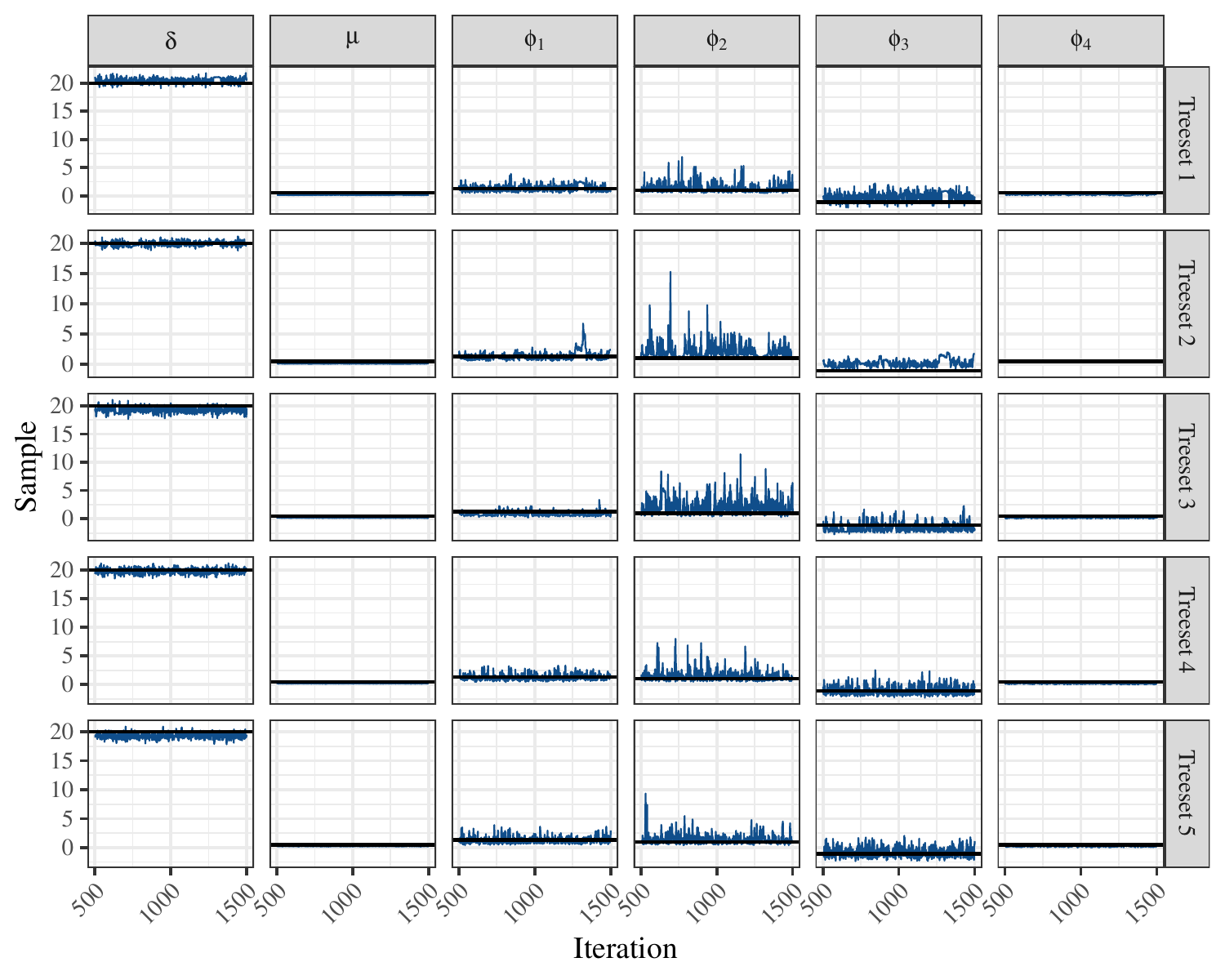}
	\caption{
		\textbf{(Incorrect sampling probability)}
		Posterior traceplots of each parameter, for each tree set.
	}
\end{figure}

\pagebreak

\subsection*{Carrying capacity}

The following parameter values and figures correspond to the simulation study described in Section \textit{\nameref{sec:sim-study-cc}}.

\subsubsection*{Ground truth parameter values}

\begin{itemize}
	\item $\phi = [2.5, 1.5, -0.1, 0.6]$
    \item $\mu = 1$
    \item $\delta = 1$
    \item $\rho = 0.1$
\end{itemize}

The type space and type change rate matrix are taken from Sections \textit{\nameref{sec:discretization}} and \textit{\nameref{sec:tc-matrix}}, respectively.

The carrying capacity is set to be $1000$.

\subsubsection*{Additional figures}

\begin{figure}[H]
	\centering
	\includegraphics[width=1\textwidth]{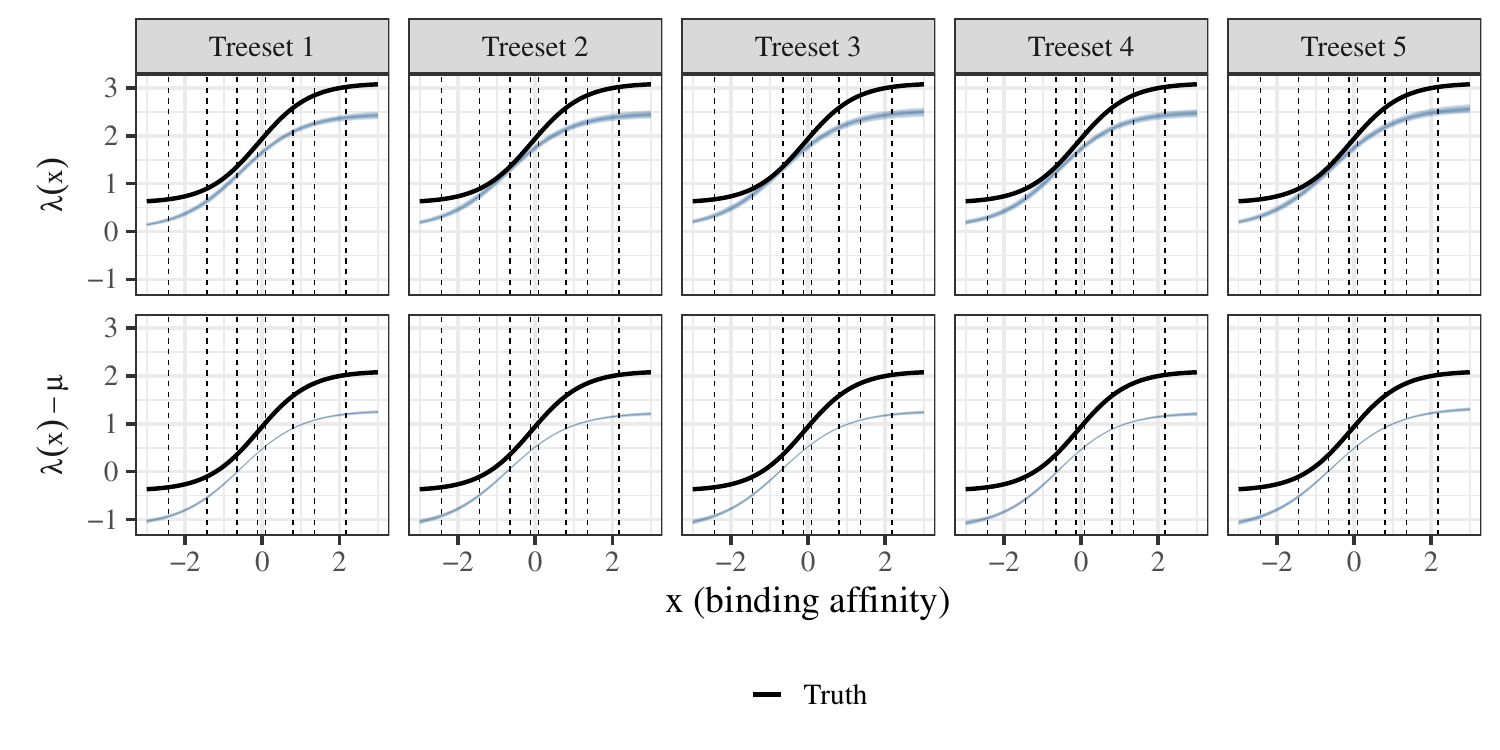}
	\caption{
		\textbf{(Carrying capacity)}
		Posteriors of the birth rate sigmoid and difference in birth rate and death rate, for each tree set.
		Shaded regions represent credible intervals (the widest being 90\%).
		The dashed lines show the values of the type space, i.e. the points at which the birth rate sigmoid is inferred.
	}
\end{figure}

\begin{figure}[H]
	\centering \small
	\includegraphics[width=1\textwidth]{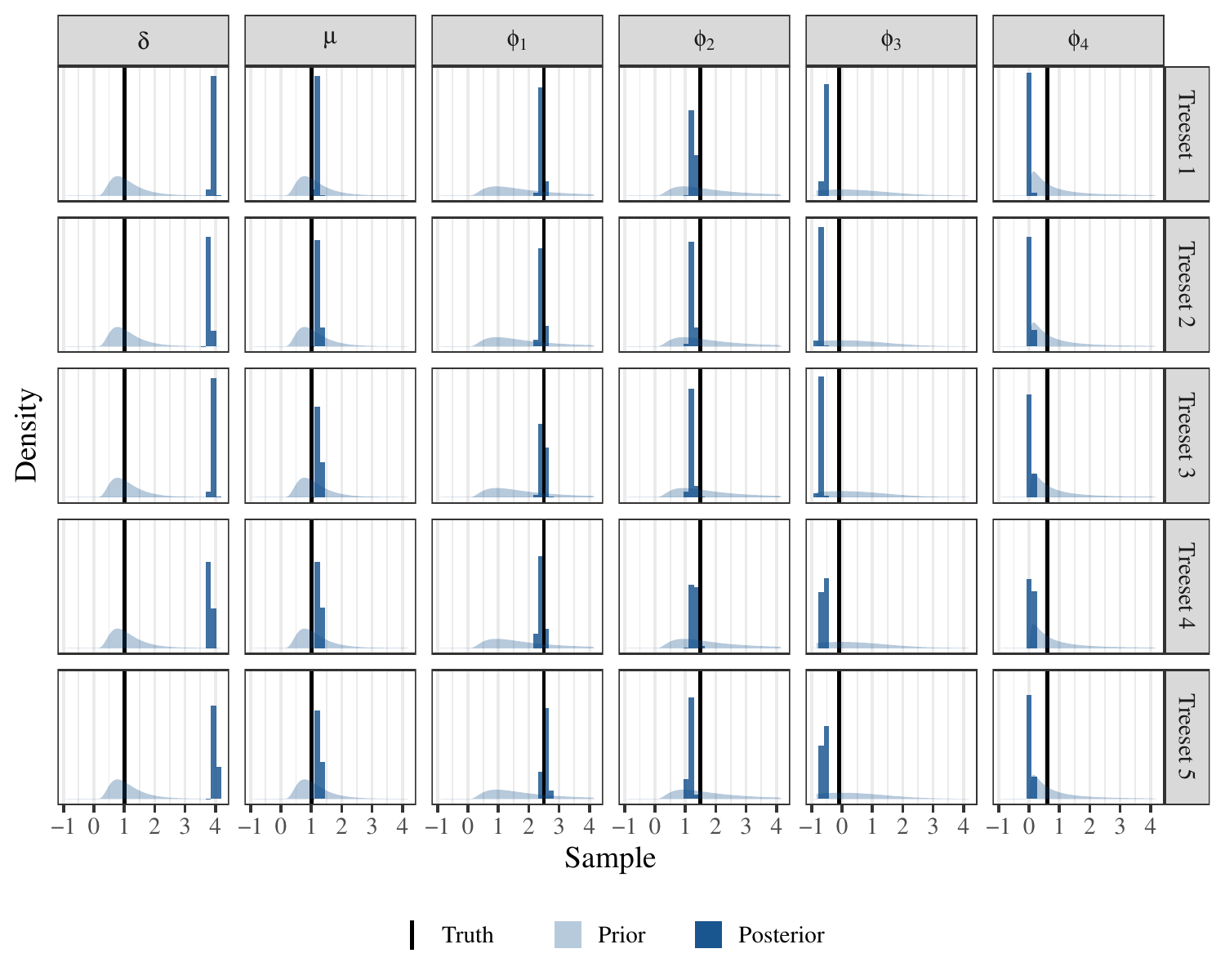}
	\caption{
		\textbf{(Carrying capacity)}
		Posterior histograms and prior density curves of each parameter, for each tree set.
	}
\end{figure}

\begin{figure}[H]
	\centering \small
	\includegraphics[width=1\textwidth]{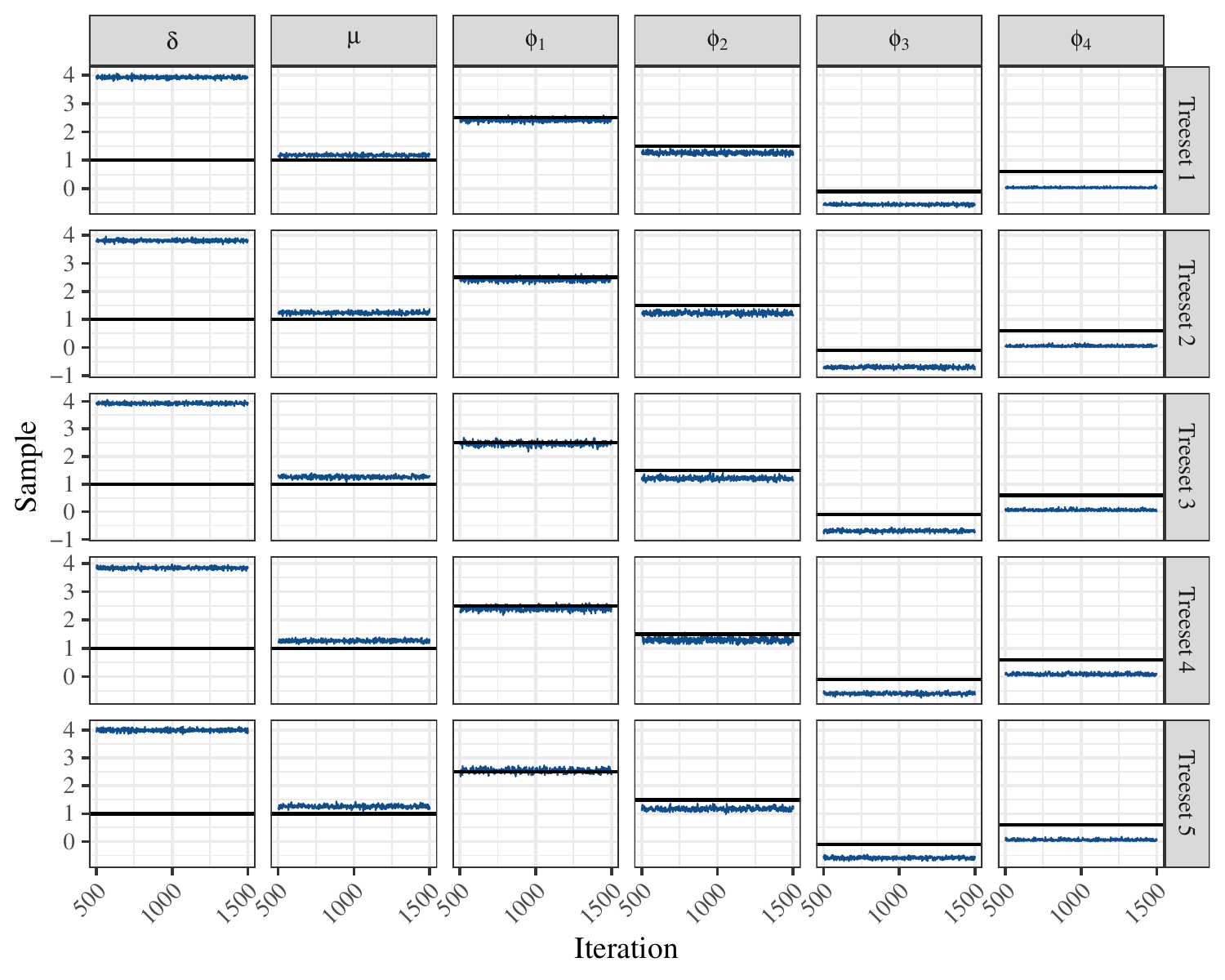}
	\caption{
		\textbf{(Carrying capacity)}
		Posterior traceplots of each parameter, for each tree set.
	}
\end{figure}

\pagebreak

\subsection*{Soft carrying capacity}

The following parameter values and figures correspond to the simulation study described in Section \textit{\nameref{sec:sim-study-cc-soft}}.

\subsubsection*{Ground truth parameter values}

\begin{itemize}
	\item $\phi = [2.5, 1.5, -0.1, 0.6]$
    \item $\mu = 1$
    \item $\delta = 1$
    \item $\rho = 0.1$
\end{itemize}

The type space and type change rate matrix are taken from Sections \textit{\nameref{sec:discretization}} and \textit{\nameref{sec:tc-matrix}}, respectively.

The carrying capacity is set to be $1000$.

\subsubsection*{Additional figures}

\begin{figure}[H]
	\centering
	\includegraphics[width=1\textwidth]{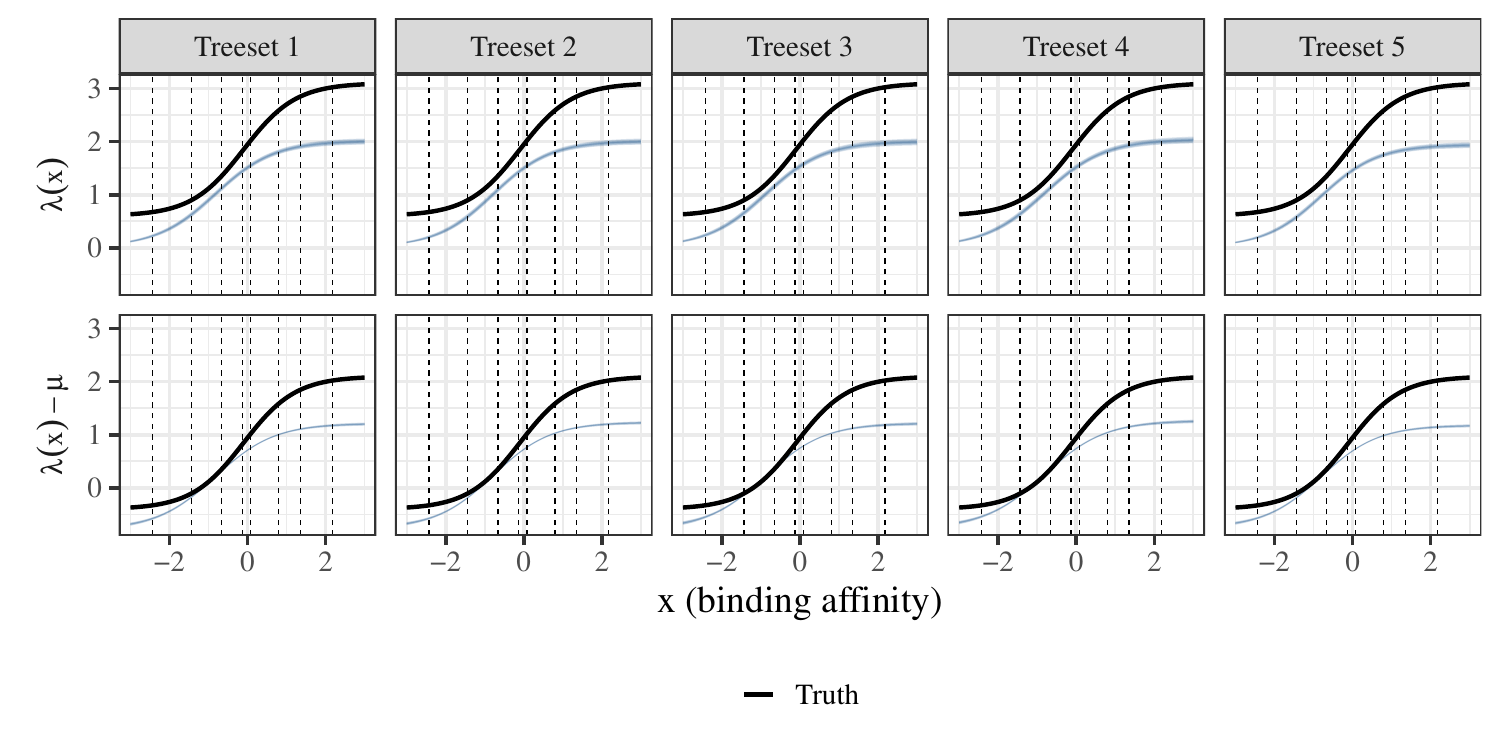}
	\caption{
		\textbf{(Soft carrying capacity)}
		Posteriors of the birth rate sigmoid and difference in birth rate and death rate, for each tree set.
		Shaded regions represent credible intervals (the widest being 90\%).
		The dashed lines show the values of the type space, i.e. the points at which the birth rate sigmoid is inferred.
	}
\end{figure}

\begin{figure}[H]
	\centering \small
	\includegraphics[width=1\textwidth]{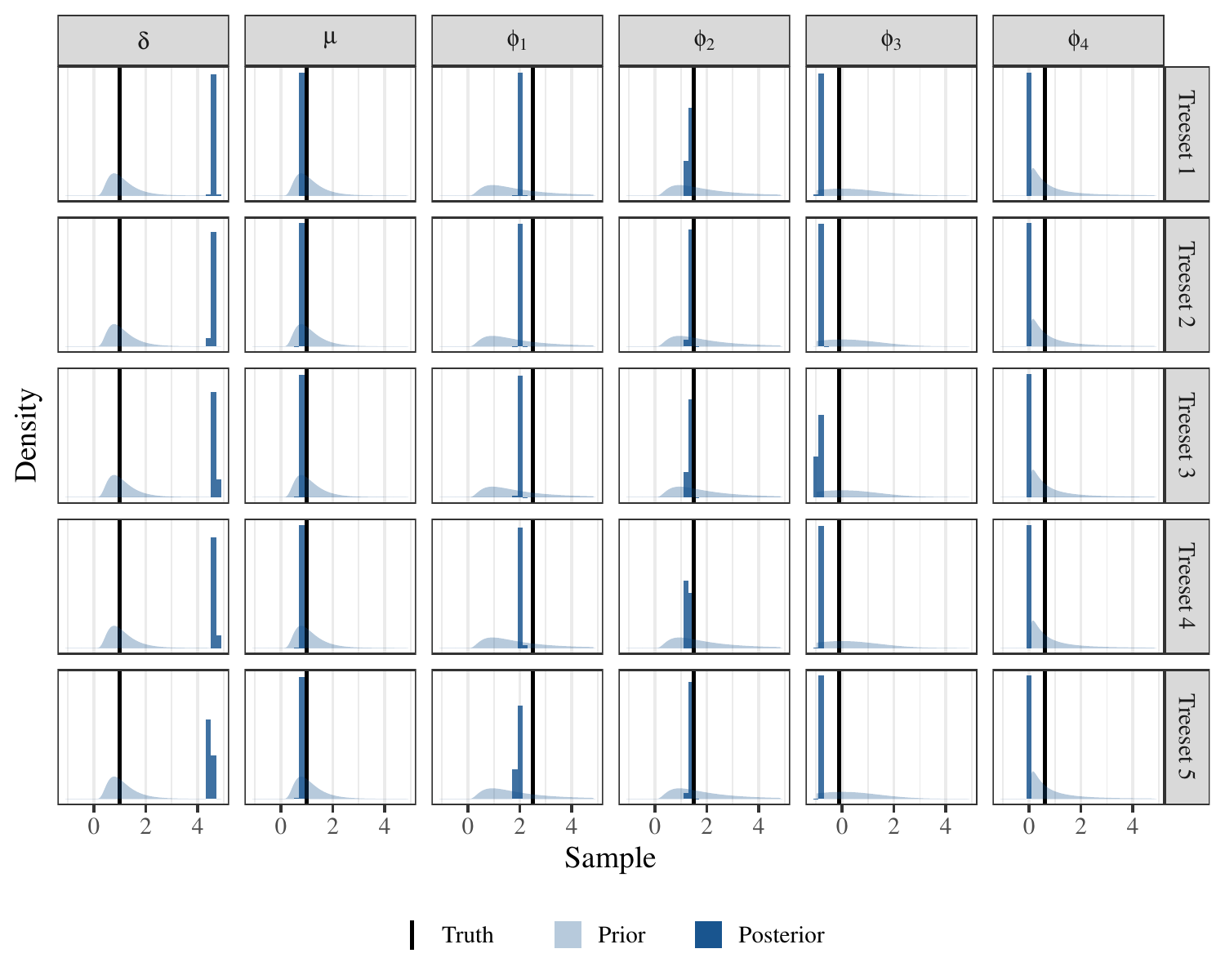}
	\caption{
		\textbf{(Soft carrying capacity)}
		Posterior histograms and prior density curves of each parameter, for each tree set.
	}
\end{figure}

\begin{figure}[H]
	\centering \small
	\includegraphics[width=1\textwidth]{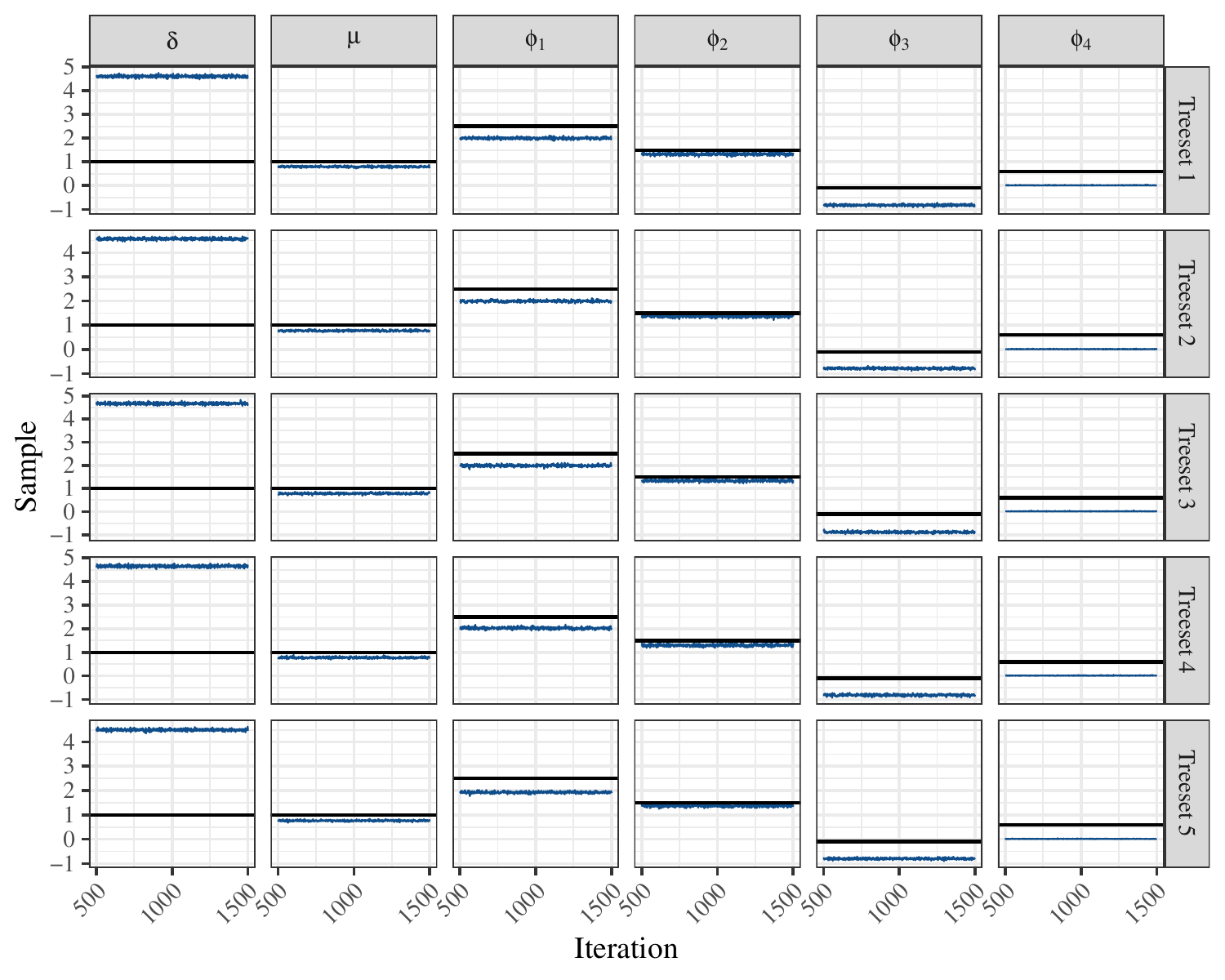}
	\caption{
		\textbf{(Soft carrying capacity)}
		Posterior traceplots of each parameter, for each tree set.
	}
\end{figure}

\pagebreak

\subsection*{Carrying capacity with sequence-level mutation process}

The following parameter values and figures correspond to the simulation study described in Section \textit{\nameref{sec:sim-study-cc-slm}}.

\subsubsection*{Ground truth parameter values}

\begin{itemize}
	\item $\phi = [2.5, 1.5, -0.1, 0.6]$
    \item $\mu = 1$
    \item $\rho = 0.1$
\end{itemize}

The carrying capacity is set to be $1000$.

\subsubsection*{Additional figures}

\begin{figure}[H]
	\centering
	\includegraphics[width=1\textwidth]{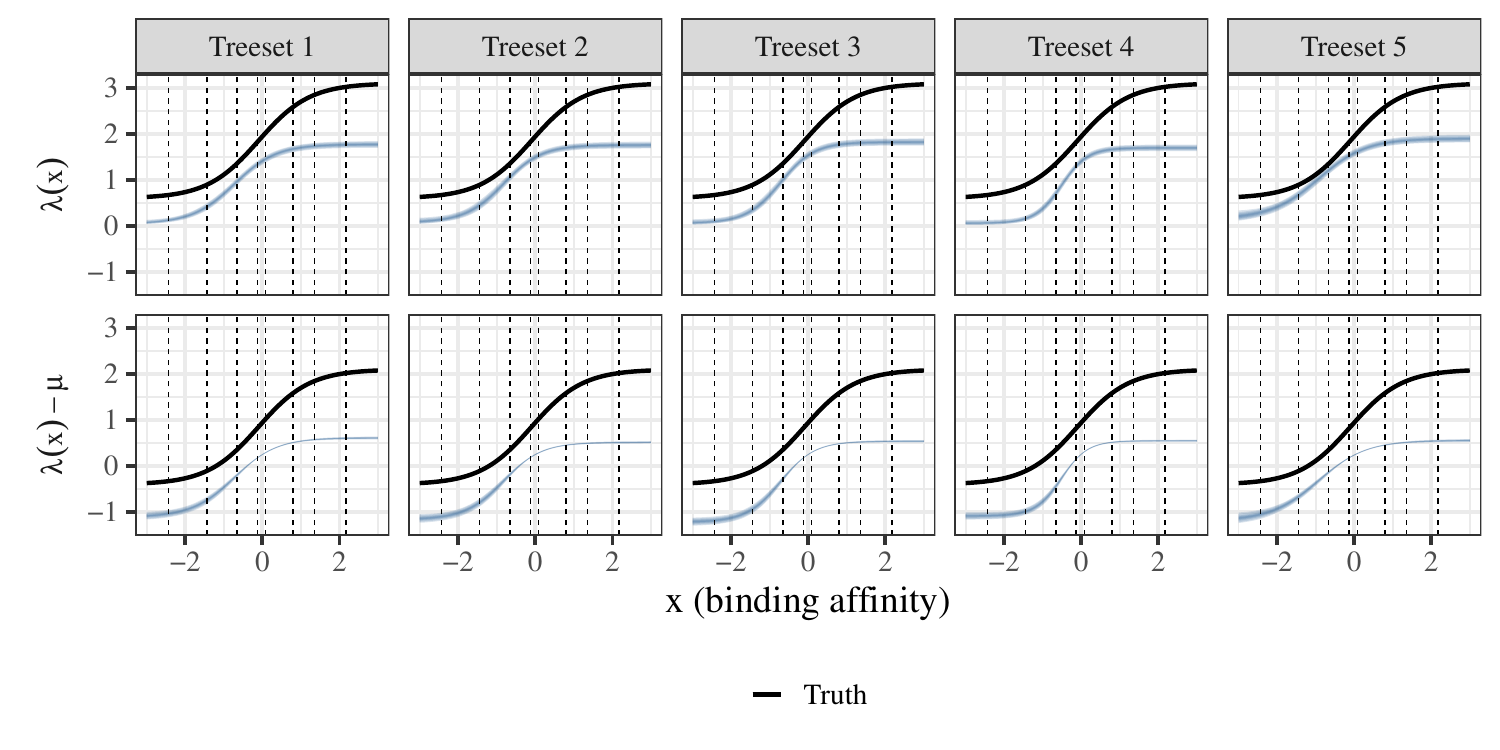}
	\caption{
		\textbf{(Carrying capacity with sequence-level mutation process)}
		Posteriors of the birth rate sigmoid and difference in birth rate and death rate, for each tree set.
		Shaded regions represent credible intervals (the widest being 90\%).
		The dashed lines show the values of the type space, i.e. the points at which the birth rate sigmoid is inferred.
	}
\end{figure}

\begin{figure}[H]
	\centering \small
	\includegraphics[width=1\textwidth]{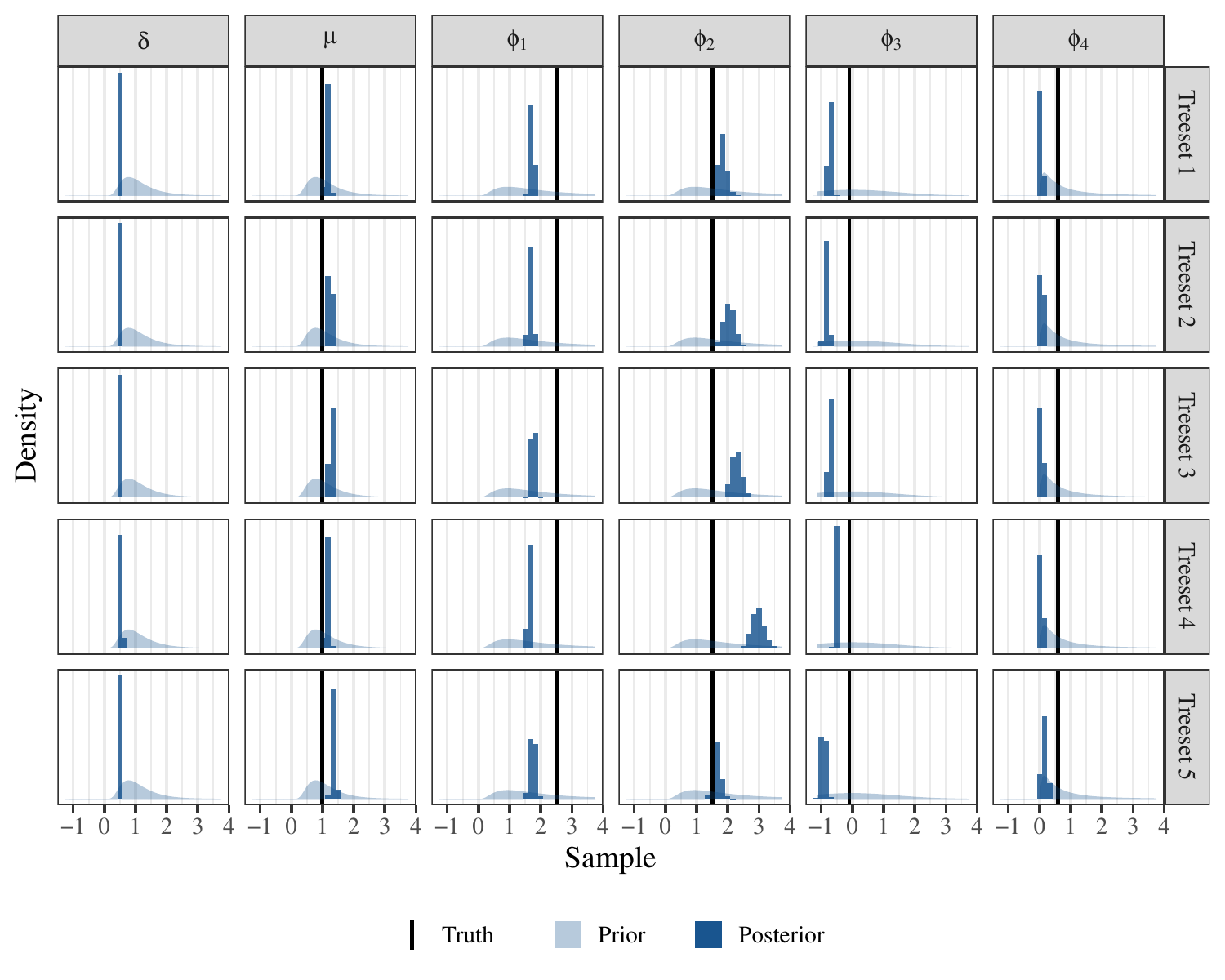}
	\caption{
		\textbf{(Carrying capacity with sequence-level mutation process)}
		Posterior histograms and prior density curves of each parameter, for each tree set.
	}
\end{figure}

\begin{figure}[H]
	\centering \small
	\includegraphics[width=1\textwidth]{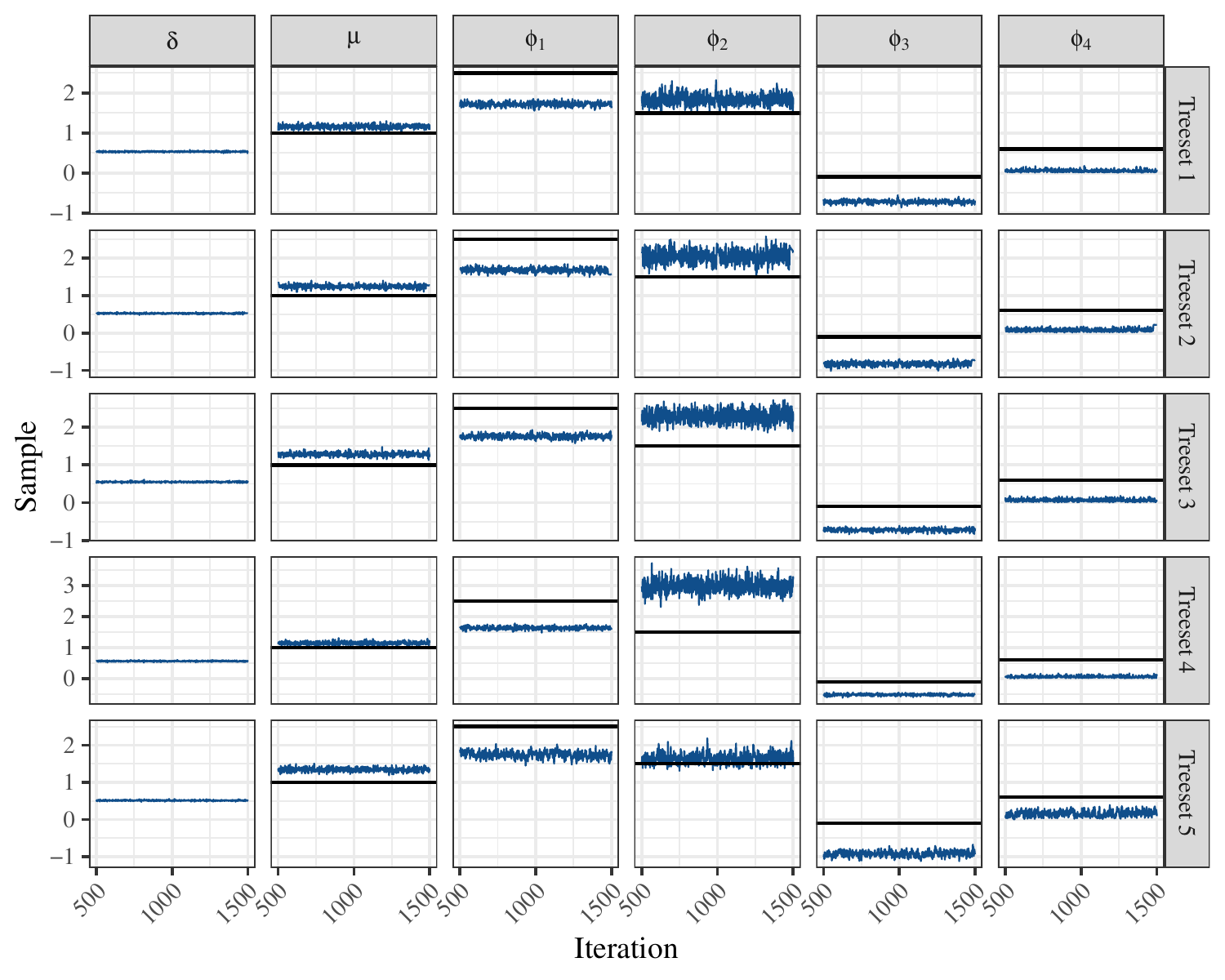}
	\caption{
		\textbf{(Carrying capacity with sequence-level mutation process)}
		Posterior traceplots of each parameter, for each tree set.
	}
\end{figure}

\pagebreak

\section*{Additional data analysis figures} \label{sec:appendix-da}

The following figures correspond to the data analysis described in Section \textit{\nameref{sec:data}}.
To better understand the effect of phylogenetic uncertainty on our inference, we inferred posteriors for five tree sets, where each tree set was created by uniformly randomly selecting one tree from the BEAST chains for each germinal center (see Section \textit{\nameref{sec:data-description}}).
The first figure visualizes the posterior over sigmoid curves inferred for each tree set.
The second figure decomposes this posterior visualization into posterior histograms for each sigmoid parameter individually, as well as for the other branching process parameters.
The third figure shows traceplots for each of these parameters.
The final figure shows diagnostics of the posterior predictive distribution obtained when combining the inferences for each tree set.

\begin{figure}[H]
	\centering
	\includegraphics[width=1\textwidth]{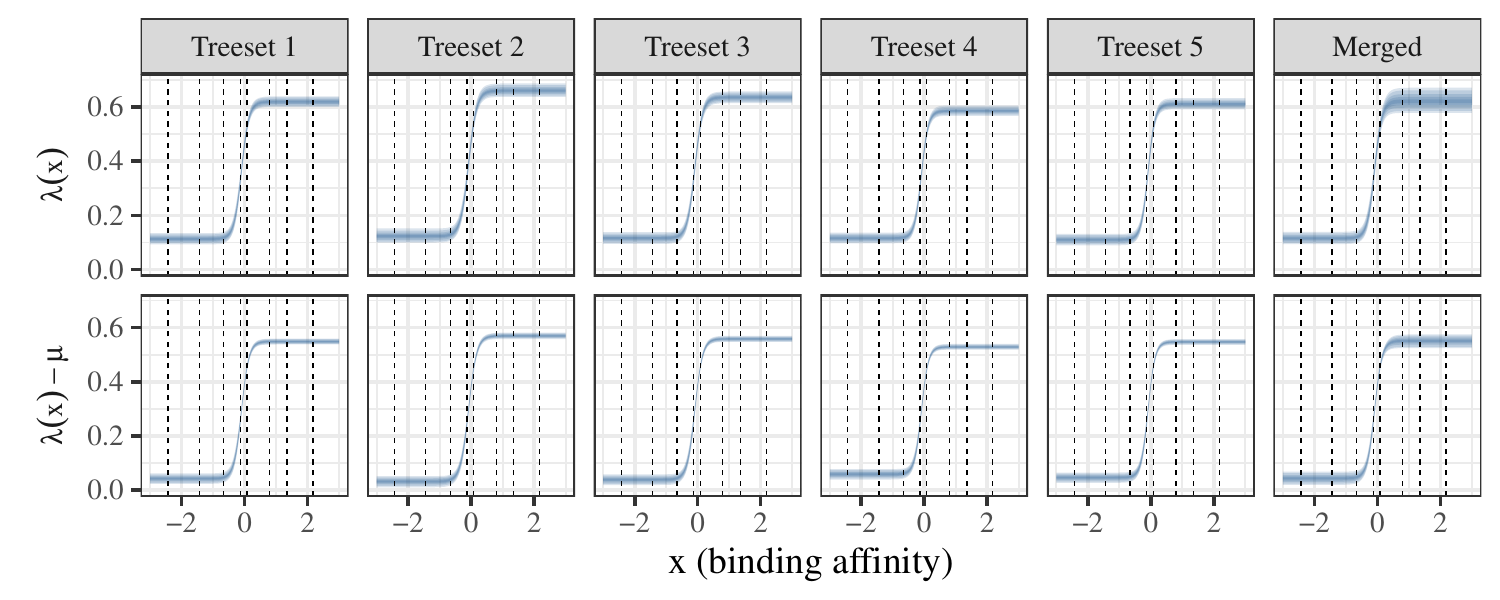}
	\caption{
		\textbf{(Data analysis)}
		Posteriors of the birth rate sigmoid and difference in birth rate and death rate, for each tree set.
		The last column displays the concatenation of the other five Markov chains, to illuminate the uncertainty we may not be capturing by using only one tree per germinal center to form our tree sets.
	}
\end{figure}

\begin{figure}[H]
	\centering \small
	\includegraphics[width=1\textwidth]{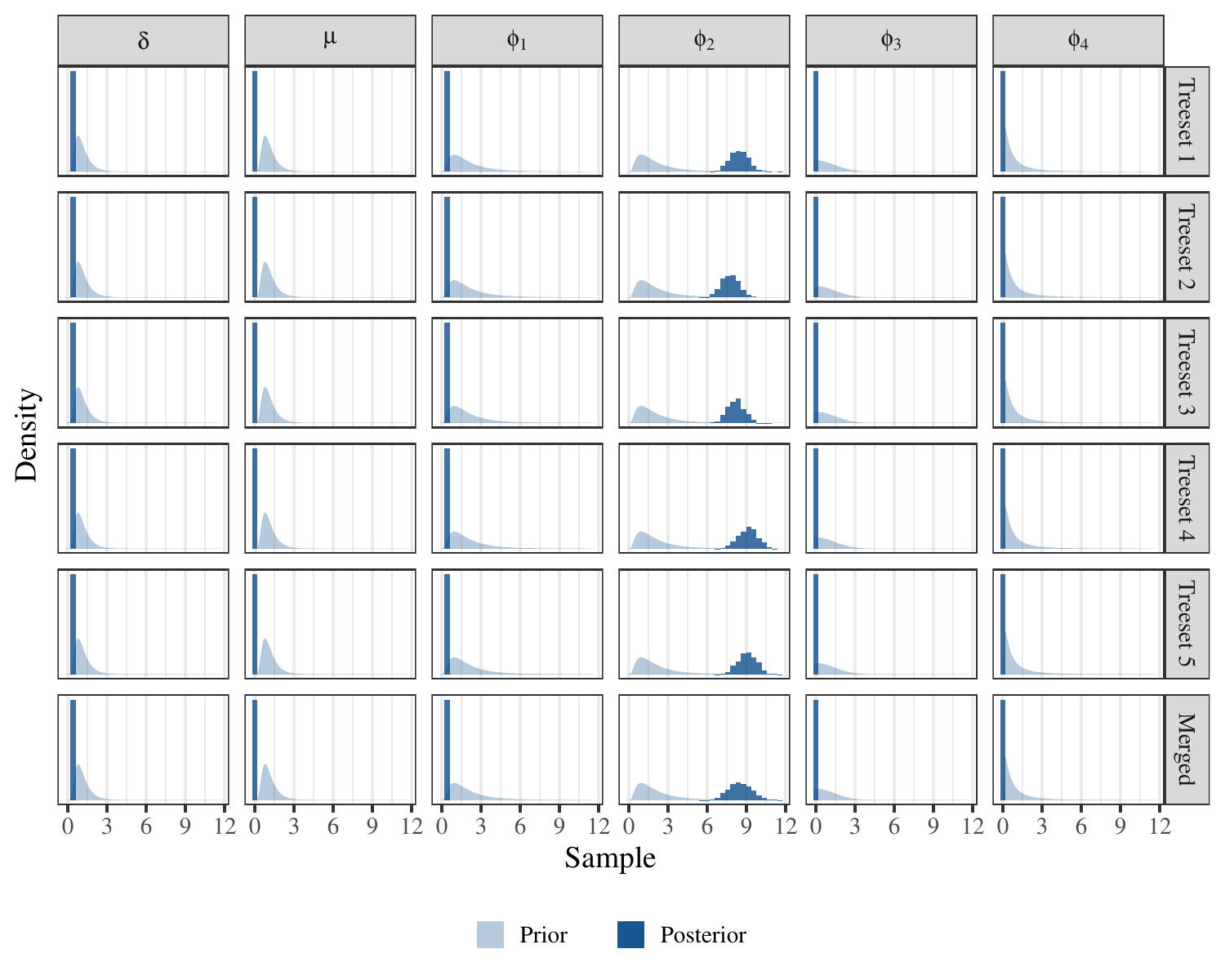}
	\caption{
		\textbf{(Data analysis)}
		Posterior histograms and prior density curves of each parameter, for each tree set.
	}
\end{figure}

\begin{figure}[H]
	\centering \small
	\includegraphics[width=1\textwidth]{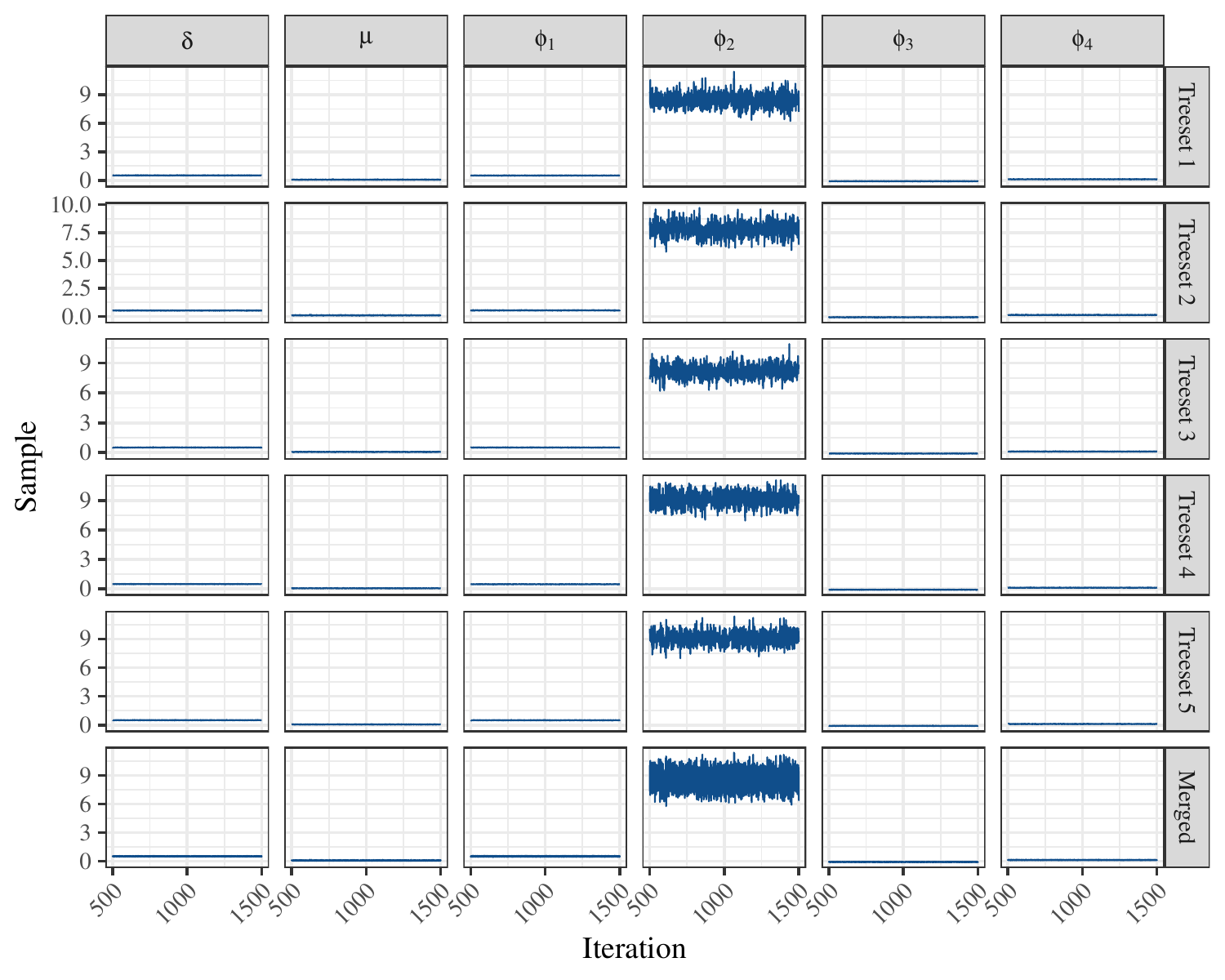}
	\caption{
		\textbf{(Data analysis)}
		Posterior traceplots of each parameter, for each tree set.
	}
\end{figure}

\begin{figure}[H]
	\centering \small
	\includegraphics[width=1\textwidth]{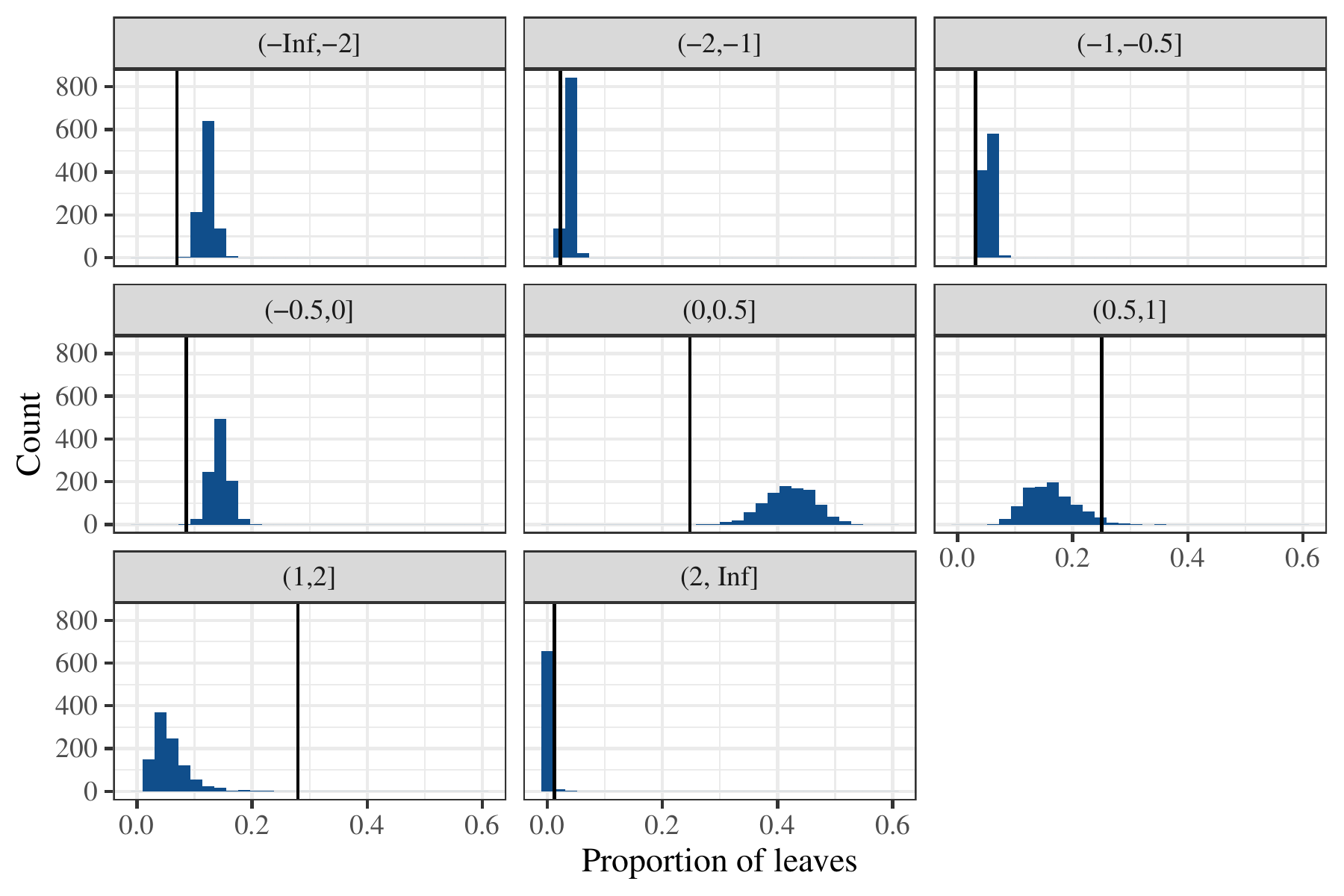}
	\caption{
		\textbf{(Data analysis)}
		Posterior predictive diagnostics.
		Histograms capture the distribution of the proportion of leaves in each affinity bin, over replicates produced by each posterior sample across all tree sets.
		Black vertical lines show the proportion of leaves with each affinity value observed in the experimental data.
	}
\end{figure}

\begin{figure}[H]
	\centering \small
	\includegraphics[width=1\textwidth]{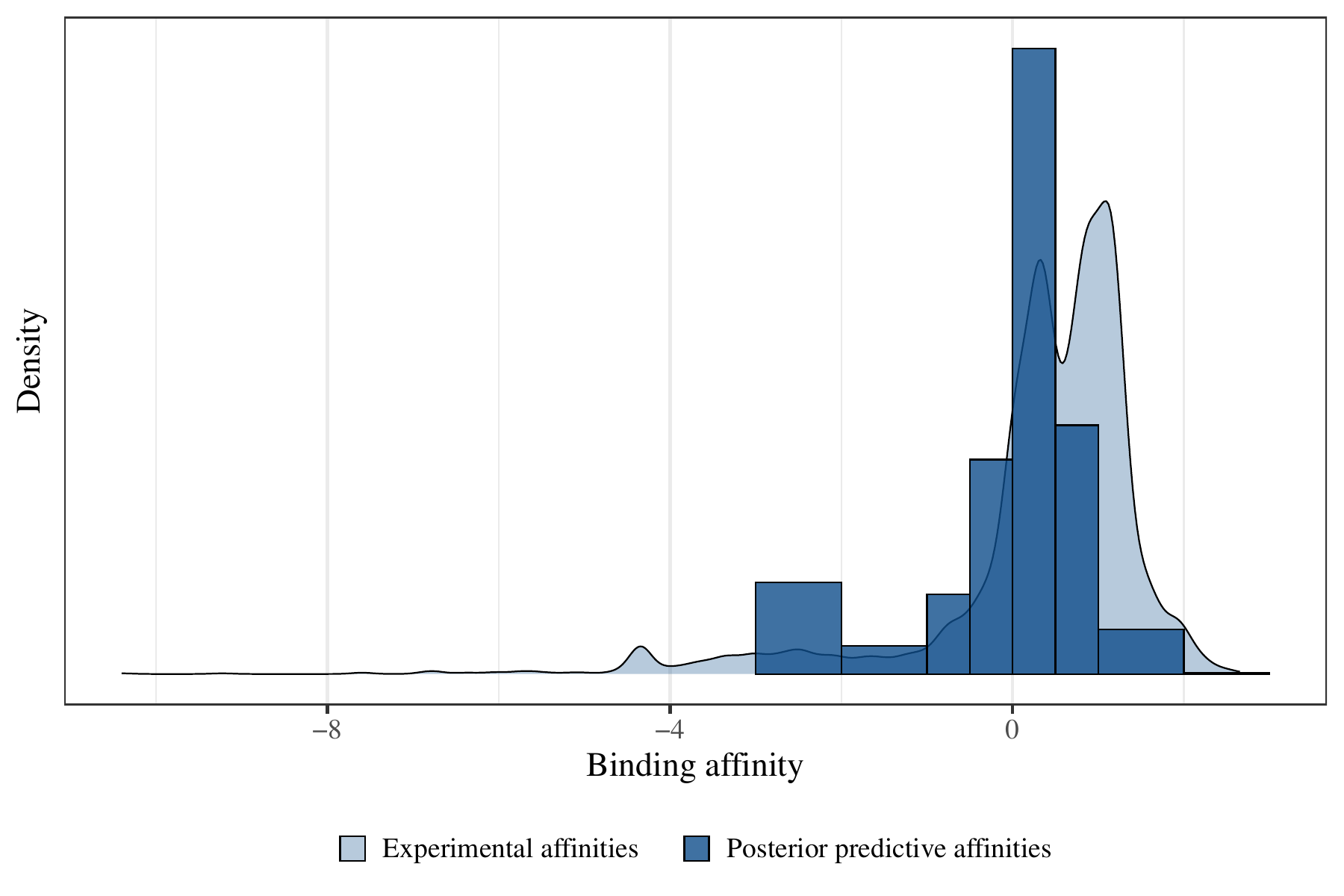}
	\caption{
		\textbf{(Data analysis)}
		Posterior predictive affinity distribution.
		The histogram captures the distribution of leaf affinities over replicates produced by each posterior sample across all tree sets.
		The density curve captures the distribution of leaf affinities observed in the experimental data.
		Note that this curve differs from the one drawn in Fig \ref{fig:type-space}, which captures both leaf and internal nodes.
	}
\end{figure}

\pagebreak

\section*{Priors} \label{sec:appendix-priors}

\begin{table}[h]
\centering
\begin{tabular}{ll}
\hline
\bf{Parameter} & \bf{Prior} \\ \thickhline
$\phi_1$ & $\text{log-normal}(0.5, 0.75)$ \\ \hline
$\phi_2$ & $\text{log-normal}(0.5, 0.75)$ \\ \hline
$\phi_3$ & $\text{normal}(0, 2)$ \\ \hline
$\phi_4$ & $\text{log-normal}(-0.5, 1.2)$ \\ \hline
$\mu$ & $\text{log-normal}(0, 0.5)$ \\ \hline
$\delta$ & $\text{log-normal}(0, 0.5)$ \\ \hline
\end{tabular}
\caption{
	Priors used in data analysis and simulation studies.
	The log-normal distributions are parameterized in terms of scale, and the normal distribution is parameterized in terms of variance.
}
\label{tab:priors}
\end{table}

\end{document}